\documentclass[11pt,fullpage]{article} 
\usepackage[centertags]{amsmath}
\usepackage{amsfonts}
\usepackage{amssymb}
\usepackage{amsthm}
\usepackage{eepic}
\usepackage{newlfont}
\usepackage{algorithm}
\usepackage{algorithmic}
\usepackage{tikz}

\theoremstyle{plain}
\newtheorem{thm}{Theorem}[section]
\newcommand{\BTHM}{\begin{thm}} \newcommand{\ETHM}{\end{thm}}
\newtheorem{cor}[thm]{Corollary}
\newcommand{\BCR}{\begin{cor}} \newcommand{\ECR}{\end{cor}}
\newtheorem{lem}[thm]{Lemma}
\newcommand{\BL}{\begin{lem}}   \newcommand{\EL}{\end{lem}}
\newtheorem{clm}[thm]{Claim}
\newcommand{\BCM}{\begin{clm}}   \newcommand{\ECM}{\end{clm}}
\newtheorem{prop}[thm]{Proposition}
\newcommand{\BP}{\begin{prop}}   \newcommand{\EP}{\end{prop}}
\newtheorem{assm}[thm]{Assumption}
\newcommand{\BASM}{\begin{assm}}   \newcommand{\EASM}{\end{assm}}

\theoremstyle{definition}
\newtheorem{defn}{Definition}[section]
\newcommand{\BD}{\begin{defn}}   \newcommand{\ED}{\end{defn}}
\newtheorem{con}[thm]{Conjecture}
\newcommand{\BCONJ}{\begin{con}}   \newcommand{\ECONJ}{\end{con}}

\theoremstyle{definition}
\newtheorem{problem}[thm]{Problem}
\newcommand{\BPR}{\begin{problem}}   \newcommand{\EPR}{\end{problem}}

\newenvironment{rem}{\noindent{\bf Remark:~~}}{}
\newcommand{\BREM}{\begin{rem}} \newcommand{\EREM}{\end{rem}}
\newenvironment{discussion}{\noindent{\bf Discussion:~~\\}}{}
\newcommand{\BDIS}{\begin{discussion}} \newcommand{\EDIS}{\end{discussion}}

\numberwithin{equation}{section}

\def\blackslug
     {\hbox{\hskip 1pt\vrule width 8pt height 8pt depth 1.5pt\hskip 1pt}}
\def\qed{\quad\blackslug\lower 8.5pt\null\par}

\newtheorem{exmp}[thm]{Example}
\newcommand{\BEX}{\begin{exmp}} \newcommand{\EEX}{\end{exmp}}
\newcommand{\BF}{\begin{fact}}   \newcommand{\EF}{\end{fact}}
\newcommand{\Bcr}{\begin{techcorr}}
\newcommand{\Ecr}{\end{techcorr}}
\newcommand{\BDS}{\begin{description}}
\newcommand{\EDS}{\end{description}}
\newcommand{\BE}{\begin{enumerate}}
\newcommand{\EE}{\end{enumerate}}
\newcommand{\BI}{\begin{itemize}}
\newcommand{\EI}{\end{itemize}}
\renewenvironment{proof}{\noindent{\bf Proof:~~}}{\qed}
\newcommand{\BPF}{\begin{proof}}
\newcommand{\EPF}{\end{proof}}

\newcommand{\BB}{\begin{enumerate}}
\newcommand{\EB}{\end{enumerate}}

\title{Target Set Selection for Conservative Populations}

\author{Uriel Feige \qquad \qquad Shimon Kogan \\ \\
  Department of Computer Science and Applied Mathematics \\
          Weizmann Institue, Rehovot 76100, Israel \\
          \ \{uriel.feige,shimon.kogan\}@weizmann.ac.il }

\begin{document}
\maketitle
\begin{abstract}
Let $G = (V,E)$ be a graph on $n$ vertices, where $d_v$ denotes the degree of vertex $v$, and $t_v$ is a threshold associated with $v$. 
We consider a process in which initially a set $S$ of vertices becomes active, and thereafter, in discrete time steps, every vertex $v$ that has at least $t_v$ active neighbors becomes active as well.
The set $S$ is {\em contagious} if eventually all $V$ becomes active. The {\em target set selection} problem TSS asks for the smallest contagious set. TSS is NP-hard and moreover, notoriously difficult to approximate.

In the {\em conservative} special case of TSS, $t_v > \frac{1}{2}d_v$ for every $v \in V$. In this special case, TSS can be approximated within a ratio of $O(\Delta)$, where $\Delta = \max_{v \in V}[d_v]$. In this work we introduce a more general class of TSS instances that we refer to as {\em conservative on average} (CoA), that satisfy the condition $\sum_{v\in V} t_v > \frac{1}{2}\sum_{v \in V} d_v$.  We design approximation algorithms for some subclasses of CoA. For example, if $t_v \geq \frac{1}{2}d_v$ for every $v \in V$,
we can find in polynomial time a contagious set of size $\tilde{O}\left(\Delta \cdot OPT^2 \right)$, where $OPT$ is the size of a smallest contagious set in $G$. 
We also provide several hardness of approximation results. For example, assuming the unique games conjecture, we prove that TSS on CoA instances with $\Delta \le 3$ cannot be approximated within any constant factor.

We also present results concerning the fixed parameter tractability of CoA TSS instances, and approximation algorithms for a related problem, that of TSS with partial incentives.
\end{abstract}



\section{Introduction}
Let $G = (V,E)$ be a graph with $n$ vertices and a threshold function $t: V(G) \rightarrow \mathbb{N}$.
We consider the following iterative process on $G$.
Initially, all vertices are {\em inactive}.
Then, a subset $S$ of vertices is selected and all vertices in $S$ become active. After that, in every iteration, for every
inactive vertex $v$, if at least $t(v)$ of its neighbors were active in the previous iteration, then $v$ becomes active. The process runs until either all vertices are active or no additional vertices can update states from inactive to active.
A subset $S$ is called a {\em contagious} if activating it results in all the vertices in $G$ becoming active in the end of the process.
Given a graph $G$ and a threshold function $t$, the problem of finding a contagious set $S$ of smallest possible size is referred to as {\em Target Set Selection} (TSS).
A $k$-approximate solution to the TSS problem returns a contagious set (or target set, we shall use these two terms interchangeably) of size at most
$k \cdot OPT$, where $OPT$ is the size of the optimal solution to TSS.

One motivation for the TSS problem is as follows. The graph $G$ may represent a social network, where the vertices may represent individual agents, and edges may represent being friends. Becoming active may correspond to adopting some novelty (a new technology, a new political agenda, a new fashion style, and so on). An agent becomes active only if sufficiently many of its friends are active (for example, the novelty involves interaction with other agents, and its attractiveness to an agent increases as more of its friends adopt it). TSS models the question of how to introduce a novelty in a cost effective way: is there some small number of ``influential agents" such that if they are convinced to adopt the novelty, the novelty will spread to the rest of the network.

The TSS problem is notoriously hard to approximate. 
It is known (folklore) that the TSS problem can be approximated within a factor of $O(n/\log n)$, and this is currently the best approximation ratio for the problem, even for graphs of bounded degree. Clearly, such a poor approximation ratio can hold only in graphs (and associated threshold functions $t$) for which the optimal target set (whose size we done by $OPT$) is very small, because getting an approximation ratio of $n/OPT$ is trivial (initially activate the whole graph).
One may ask whether there are natural classes of TSS instances for which there are much better approximation algorithms. We now describe one such class.

We may say that an agent is {\em conservative} if it becomes active only if a strict majority of its neighbors are active. Intuitively, if all agents are conservative, then target sets need to be large, and TSS can be approximated within improved factors. Indeed, for bounded degree graphs, this is the case.
Let $d(v)$ denote the degree of vertex $v$ in graph $G$, and let $\Delta$ denote the maximum degree in $G$. A threshold function is referred to as a {\em strict majority threshold function} if for every vertex $v$ it holds that $t(v) > \frac{d(v)}{2}$.

\BTHM
\label{thm:strictMajority}
For instances with strict majority threshold functions, there are approximation algorithms with approximation ratio $O(\Delta)$. Moreover, in the absence of a dependency on $n$, having a dependency on $\Delta$ in the approximation ratio is unavoidable.
\ETHM

The proof of Theorem~\ref{thm:strictMajority} is fairly easy and is omitted here. We just remark that
the algorithmic result of Theorem~\ref{thm:strictMajority} follows from the results of~\cite{DBLP:journals/tcs/AckermanBW10}, whereas the hardness result can be proved as in the proof of Theorem~\ref{degenerate_colorally13_main}.

We believe that the assumption that agents are conservative is natural in many situations. However, the assumption that {\em all} agents are conservative might be too strong. Hence in this work, we consider instances in which agents are conservative {\em on average}. There is more than one way in which the term ``conservative on average" can be defined, and we choose to use the following definition.

\BD
\label{def:conservative}
Consider an instance of TSS with $n$ vertices (also referred to as agents), in which for every vertex  $i$, $d_i$ denotes its degree and $t_i$ denotes its threshold. We say that agents are {\em conservative on average} (CoA) if $\sum_{i=1}^n t_i \ge \frac{1}{2}\sum_{i=1}^n d_i$.
\ED

Observe that in our definition we introduced the condition $\sum_{i=1}^n t_i \ge \frac{1}{2}\sum_{i=1}^n d_i$ (with weak inequality) rather than $\sum_{i=1}^n t_i > \frac{1}{2}\sum_{i=1}^n d_i$ (with strict inequality, as was done in the notion of ``conservative" for individual agents). This is done for convenience. Its effect on the results is negligible, because given any instance of TSS that satisfies the weaker definition, we can raise by one the threshold of one of the vertices, and get an instance that satisfies the strong definition. This change can change the size of the minimum target set by at most~1, which is negligible compared to other approximation ratios that appear in our results.

An alternative definition that we considered for being conservative on average is that of $\sum_{i=1}^n \frac{t_i}{d_i} \ge \frac{n}{2}$. (The alternative definition says that the average value of $\frac{t_i}{d_i}$ is at least $\frac{1}{2}$. In contrast, Definition~\ref{def:conservative} is equivalent to requiring that the weighted average is at least $\frac{1}{2}$, where the weight of a vertex is its degree.) However, TSS instances that satisfy this alternative definition are not easier to approximate than general TSS instances. This is because given any instance of TSS, we can add to it $n$ vertices of degree~1 and threshold~1, all connected to one of the vertices of the original instance. The size of the minimum target set does not change (none of the new vertices are in the minimum target set), but the new instance does satisfy  $\sum_{i=1}^{2n} \frac{t_i}{d_i} \ge \frac{2n}{2}$ (here $2n$ is the new number of vertices, replacing the original $n$).

Returning to Definition~\ref{def:conservative},
it is not hard to see that instances of TSS that are CoA may have very small target sets, even if they have bounded degree. A simple example is a cycle in which all thresholds are~1, as there any single vertex activates the whole graph. This makes the design of approximation algorithms for such instances considerably more difficult compared to the case in which all agents are conservative. Indeed, in the current work we are not able to determine whether CoA TSS instances can be approximated within approximation ratios that are significantly better than those for general TSS instances. This remains as an interesting open question. What we are able to do is to design approximation algorithms for some natural subclasses of CoA, and also to provide some new hardness of approximation results that apply in CoA settings.

We also consider a related problem, that of TSS with partial incentives \cite{DBLP:conf/sirocco/CordascoGRV15}, that we denote by $TSS_P$. An instance of $TSS_P$ is similar to an instance of TSS, and one seeks the minimum value for $\sum_{i=1}^n p_i$ (where $p_i \ge 0$ for all $i$) such that in the instance with thresholds $t_i - p_i$, if all agents with threshold~0 are activated, the whole graph becomes active.


Before proceeding to describe our new results and previous results in more detail, we introduce some notation that is hopefully intuitive. Here are some representative examples for the use of our notation:
\begin{itemize}
\item TSS($d=c$, $t=q$) denotes the class of instances of TSS in which the $G$ is $c$-regular and all thresholds have value $q$.
\item TSS($d \leq c$, $t \in \{1,2\}$): instances in which $G$ is of maximum degree $c$ and every threshold is either~1 or~2.
\item TSS($t_v \geq \frac{1}{2} d_v$): instances where for each vertex the threshold is at least half of its degree.
\end{itemize}

Throughout this paper, $n$ represents the number of vertices in the input graph $G(V,E)$, $\Delta$ represents its maximum degree, $\bar{d}$ represents its average degree, and $OPT$ represents the value of the optimal solution (which is the size of the smallest contagious set).

\subsection{Main results}

A natural class of instances within CoA is one in which for every agent $i$ the threshold $t_i$ satisfies the weak majority condition $t_i \ge \frac{d_i}{2}$ (instead of the strong majority condition, which makes the agent conservative).

\BTHM\label{alg_mainthm_majority1_statement_main}
Let $(G,t)$ be an instance of  TSS($t_v \geq \frac{1}{2}d_v$).
Then a contagious set of size
$O\left(\Delta \cdot OPT^2  \cdot \sqrt{\log(\Delta \cdot OPT)} \right)$ can be found in polynomial time.
\ETHM

Observe that unlike the case of Theorem~\ref{thm:strictMajority}, in Theorem~\ref{alg_mainthm_majority1_statement_main} the approximation ratio depends not only on $\Delta$,  but also on $OPT$. We do not know whether this dependency on $OPT$ is necessary. However, we do know that even with this dependency on $OPT$, the approximation ratio of Theorem~\ref{alg_mainthm_majority1_statement_main} is better than the one that can be achieved (under reasonable complexity assumptions) for general instances of TSS. This follows from the following theorem.

\BTHM\label{cubicthmtargetset1OPT}
The TSS($d = 3$,$t \in \{1,2\}$) problem cannot be approximated within the ratio $f(OPT)$ for any computable function $f$, unless $FTP = W[P]$.
\ETHM


We now turn to a more demanding subclass of CoA instances. For a graph  $G = (V,E)$, a thershold function  $t: V(G) \rightarrow \mathbb{N}$ is {\em balanced} if every edge $(u,v)\in E$ satisfies  $\frac{t_u}{d_u} + \frac{t_v}{d_v} \geq 1$.
We denote the associated target set selection problem by TSS($\frac{t_u}{d_u} + \frac{t_v}{d_v} \geq 1$).
Clearly, every weak majority threshold function is balanced. Also, every balanced threshold function is conservative on average due to the following chain of inequalities:
$$\frac{1}{2}\sum_v d_v = |E| \le  \sum_{(u,v) \in E} \left(\frac{t_u}{d_u} + \frac{t_v}{d_v}\right) = \sum_{v \in V} d_v \frac{t_v}{d_v} = \sum_{v \in V} t_v$$

\BTHM\label{alg_mainthm_balanced1_statement_main}
Let $(G,t)$ be an instance of TSS($\frac{t_u}{d_u} + \frac{t_v}{d_v} \geq 1$).
Then a target set of size $O\left((\Delta \cdot OPT)^2  \cdot \sqrt{\log(\Delta \cdot OPT)} \right)$
can be found in polynomial time.
\ETHM

Another subclass of CoA, incomparable to the one with balanced threshold functions, is the following. Let $G = (V,E)$ be a graph on $n$ vertices, and let $t$ be a threshold function
$t: V(G) \rightarrow \mathbb{N}$. We say that $t$ is {\em degenerate} if in every induced subgraph $G'$ of graph $G$, there is a vertex $v$ in $G'$ such that
$t(v) \geq d_{G'}(v)$, where
$d_{G'}(v)$ is the degree of $v$ in $G'$ and $t(v)$ is the threshold of $v$ in graph $G$ (which is always identical to the threshold of $v$ in graph $G'$).
We denote the associated target set selection problem by TSS(degenerate).
\BTHM
\label{thm:degenerate}
Let $(G,t)$ be an instance of TSS(degenerate), where $t_{\max}$ is the maximum threshold of function $t$. 
Then a target set of size $t_{\max} \cdot OPT$ can be found in polynomial time.
\ETHM

Theorem~\ref{thm:degenerate} implies that TSS(degenerate) in graphs of bounded degree can be approximated within a constant factor. In general graphs, this is no longer true, as shown by the following theorem.

\BTHM\label{degenerate_colorally13_main}
The TSS(degenerate) problem cannot be approximated within the ratio $O\left(2^{\log^{1-\epsilon} n}\right)$ for any fixed constant $\epsilon>0$, unless $P=NP$.
\ETHM

We are not able to present in this work improved approximation ratios that apply to all CoA instances. However, we are able to provide a hardness result that shows that the algorithmic result of Theorem~\ref{thm:strictMajority} does not extent to CoA instances. Assuming the unique games conjecture ($UGC$, see \cite{DBLP:conf/stoc/Khot02a}), we show that even in bounded degree graphs, CoA instances cannot be approximated within any constant factor. In the following theorem and elsewhere, $\bar{d}$ denotes the average degree of a graph.

\BTHM\label{avgdegthm1_main}
The TSS($\bar{d}=4$,$d \leq 5$,$t=2$) problem is $NP$-hard to approximate within any
constant factor, assuming the unique games conjecture.
\ETHM
In contrast to Theorem~\ref{avgdegthm1_main}, for every fixed $\epsilon>0$, the TSS(${\bar{d} \leq 4-\epsilon}$,${t=2}$) problem can be approximated within a constant factor
(which depends on $\epsilon$) in polynomial time (this follows in a straightforward manner from \cite{DBLP:journals/tcs/AckermanBW10}).

\subsection{Additional results}

In our work we prove additional hardness results, some of which we describe now.
In \cite{chen1} it is proven that there is a constant $c$ such that the TSS($d \leq c$,$t \in \{1,2\}$) problem cannot be approximated within the ratio $O\left(2^{\log^{1-\epsilon} n}\right)$
for any fixed constant $\epsilon>0$, unless $NP \subseteq DTIME(n^{polylog(n)})$.
We prove the following strengthening of this result.
\BTHM\label{cubicthmtargetset1_main}
The TSS($d = 3$,$t \in \{1,2\}$) problem cannot be approximated within the ratio $O\left(2^{\log^{1-\epsilon} n}\right)$
for any fixed constant $\epsilon>0$, unless $P=NP$.
\ETHM
We note that both TSS($d \leq 3$,$t =1$) and TSS($d \leq 3$,$t =2$) can be solved in polynomial time. The first of these results is trivial, whereas, for the second result see either  \cite{DBLP:journals/ieicet/TakaokaU15} or \cite{DBLP:journals/dmtcs/KynclLV17}.

It is proven in \cite{DBLP:journals/dmtcs/KynclLV17} that the TSS($d \leq 4$,$t =2 $) is $NP$-hard. We prove the following strengthening.

\BTHM\label{tssdeg4thm2_main}
The TSS($d = 4$,$t =2$) problem is $APX$-hard.
\ETHM

It is well known that the feedback vertex set problem is equivalent to the problem TSS($t_v = d_v-1$).
In \cite{DBLP:journals/algorithmica/Rizzi09} it is proven that the feedback vertex set problem is $APX$-hard for graphs of maximum degree $4$.
We prove the following strengthening of the theorem above which might be of independent interest.

\BTHM\label{feedbackvertexset_4regular_main}
The TSS($d=4$,$t=3$) problem is $APX$-hard (that is feedback vertex set is $APX$-hard on $4$-regular graphs).
\ETHM

Another aspect that we address in our work is the fixed parameter tractability of subclasses of CoA TSS instances. In \cite{DBLP:journals/computability/BazganCNS14} it is proven that TSS($t_v \geq \frac{1}{2}d_v$) is $W[P]$-hard with respect to the parameter $OPT$. Adding also $\Delta$ as a parameter, we obtain positive results.

\BTHM\label{alg_mainthm_majority2_statement_main}
Let $(G,t)$ be an instance of  TSS($t_v \geq \frac{1}{2}d_v$).
Then one can solve TSS($t_v \geq \frac{1}{2}d_v$) in  time $2^{O(\Delta \cdot OPT \log^2 (\Delta \cdot OPT))} \cdot n^{O(1)}$.
\ETHM


\BTHM\label{alg_mainthm_balanced2_statement_main}
Let $(G,t)$ be an instance of  TSS($\frac{t_u}{d_u} + \frac{t_v}{d_v} \geq 1$).
Then one can solve TSS($\frac{t_u}{d_u} + \frac{t_v}{d_v} \geq 1$) in  time  $2^{O(\Delta^2 \cdot OPT \log^2 (\Delta \cdot OPT))} \cdot n^{O(1)}$.
\ETHM

We also consider a related problem, that of TSS with partial incentives \cite{DBLP:conf/sirocco/CordascoGRV15}, that we denote by $TSS_P$. An instance of $TSS_P$ is similar to an instance of TSS, and one seeks the minimum value for $\sum_{i=1}^n p_i$ (where $p_i \ge 0$ for all $i$) such that in the instance with thresholds $t_i - p_i$, if all agents with threshold~0 are activated, the whole graph becomes active. See Section \ref{partial_incentives_model_Sect} for a more detailed definition.

We remark that for every given instance, the optimal value of the corresponding TSS and $TSS_P$ problems differ by a factor of at most $\Delta$. Hence in bounded degree graphs the approximation ratios of these two problems differ by at most a constant factor. But still, there are significant differences in the approximability of these two problems.

$TSS(t_v = d_v)$ is NP-hard (equivalent to vertex cover). In contrast, it is shown in~\cite{cordasco2018whom} that TSS$_P$($t_v = d_v$) can be solved in polynomial time. We provide several stronger results.

\BTHM\label{surprising_polytime1}
The  TSS$_P$(degenrate) problem can be solved in polynomial time. The same applies to the $TSS_P(t_v \in \{d_v -1 , d_v\})$ problem and the $TSS_P(t_v \in \{1 , d_v\})$ problem.
\ETHM

The value of the optimal solution of $TSS_P$ instances is between~1 and $O(n^2)$. Hence obtaining an $O(n^2)$ approximation ratio is trivial. It turns out that obtaining an approximation ratio of $O(n)$ is not difficult. See Theorem~\ref{simple_pit1}. We show that for a subclass of CoA (that of weak majority thresholds), better approximation ratios are achievable.

\BTHM
\label{thm:Pweak}
Given an instance of $(G,t)$ of the TSS$_P$($t_v \geq \frac{1}{2}d_v$) problem,
an $\tilde{O}(\sqrt{n})$-approximate solution can be found in polynomial time. Furthermore the problem is $APX$-hard.
\ETHM

We stress that the approximation ratio of Theorem~\ref{thm:Pweak} for TSS$_P$($t_v \geq \frac{1}{2}d_v$) holds regardless of the value of $\Delta$, whereas similar approximation ratios are not known for TSS($t_v \geq \frac{1}{2}d_v$) when $\Delta$ is unbounded.

For strong majority thresholds, we show that even better approximation ratios can be achieved.

\BTHM
\label{thm:Pstrong}
Given an instance of ($G$,$t$) of the TSS$_P$($t_v > \frac{1}{2}d_v$) problem,
an $\tilde{O}(n^{\frac{1}{3}})$-approximate solution can be found in polynomial time.
\ETHM





\subsection{Related work}

It is known (folklore) that the TSS problem can be approximated within a factor of $O(n/\log n)$ and this is currently the best approximation ratio for the problem, even for graphs of bounded degree.

The $TSS(t=1)$ problem  can be solved in polynomial time by activating one vertex in each connected component.
The $TSS(t_v=d_v)$ problem is equivalent to the vertex cover problem as shown in \cite{chen1}, and hence can be approximated within a ratio of~2.
The $TSS(t_v=d_v-1)$ problem is equivalent to the feedback vertex set problem (follows from \cite{DBLP:journals/dam/DreyerR09}), and hence it too can be approximated within a ratio of~2. Moreover,  TSS($d \leq 3$,$t =2 $) can be solved in polynomial time (see \cite{DBLP:journals/ieicet/TakaokaU15} and \cite{DBLP:journals/dmtcs/KynclLV17}).
More generally the $TSS(t_v=d_v-k)$ problem is equivalent to the problem of deleting the minimum number of vertices in a graph such that the resulting graph is $k$-degenerate.

It follows from the results in \cite{DBLP:journals/tcs/AckermanBW10} that the TSS($t_v > \frac{1}{2}d_v$) problem (the strict majority target set selection problem)
can be approximated within a factor $O(\Delta)$ where $\Delta$ is the maximum degree of the graph.

In \cite{DBLP:journals/corr/abs-1805-10086} it is shown that the TSS problem can be approximated efficiently in graphs of bounded treewidth in polynomial time. Namely,
they prove the following theorem.
\BTHM\label{treewidth_theorem_important_related_works}
Let $(G,t)$ be an instance of $TSS$. Given a tree-decomposition of graph $G$ of width $w$, a target set of size $(w+1)OPT$ can be found in polynomial time,
where $OPT$ is the size of an optimal target set of the TSS instance.
\ETHM

Furthermore in \cite{DBLP:journals/disopt/Ben-ZwiHLN11} it is proven that the TSS problem can be solved efficiently in graphs of bounded treewidth (but the running time here depends on the treewidth).
Combining this with algorithms for finding a tree decomposition \cite{DBLP:journals/corr/abs-1304-6321} gives the following.
\BTHM\label{FPT_result_1}
Let $(G,t)$ be an instance of $TSS$. Then one can solve the instance
in  time $\Delta^{O(w \log w)} \cdot n^{O(1)}$ ,
where $w$ is the treewidth of graph $G$ and $\Delta$ the maximum degree of graph $G$.
\ETHM

In \cite{chen1} it is proven that there is a constant $c$ such that the TSS($d \leq c$,$t \in \{1,2\}$) problem cannot be approximated within the ratio $O\left(2^{\log^{1-\epsilon} n}\right)$
for any fixed constant $\epsilon>0$, unless $NP \subseteq DTIME(n^{polylog(n)})$. In \cite{DBLP:conf/approx/CharikarNW16} it is shown that assuming a conjecture on the hardness of
Planted Dense Subgraph it is impossible to approximate $TSS$ within a factor of $O(n^{\frac{1}{2}-\epsilon})$.

The following theorem is proven in \cite{DBLP:journals/computability/BazganCNS14} .
\BTHM\label{related_works_thm111}
 The TSS($t \in \{1,2\}$) and TSS($t_v \geq \frac{1}{2}d_v$) problems cannot be approximated within the ratio $f(OPT)$ for any computable function $f$, unless $FTP = W[P]$.
\ETHM





\subsection{Proof techniques}

A recurrent idea in our algorithms for the CoA TSS problems (theorems~\ref{alg_mainthm_majority1_statement_main} and~\ref{alg_mainthm_balanced1_statement_main}), is to prove that having a combination of low $OPT$ value and low maximum degree $\Delta$ implies that the graph of the instance has small treewidth (in fact, even small cutwidth). Thereafter we use Theorem~\ref{treewidth_theorem_important_related_works} as a blackbox in order to approximate the optimal target set efficiently. This approach has some similarity with an approach referred to as {\em bidimensionality} (see for example \cite{DBLP:journals/cj/DemaineH08}) which
works for problems in which low $OPT$ implies having no large grid minor, and consequently small treewidth. We cannot use the existing bidimensionality theory directly
in this paper, as contraction of edges in a $TSS$ instance does not preserve existing target sets in a meaningful way.

In algorithmic results for the partial incentive model of TSS we show that low $OPT$ implies small cutwidth of the graph, and hence also small treewidth. However, in the partial incentives model there is no known theorem similar to Theorem~\ref{treewidth_theorem_important_related_works} that we can use as a blackbox, and in fact it is not clear whether small treewidth implies a good approximation ratio. Instead, we design approximation algorithms that make direct use of the small cutwidth. 

The above algorithmic framework fails for some CoA instances. For example, consider a graph composed of a path $P_1$ of length $n/2$, a path $P_2$ of length $n/2$, and a random perfect matching between the vertices of $P_1$ and $P_2$. Such a graph has linear treewidth (with high probability). Setting $t_i = 1$ for every vertex $i \in P_1$ and $t_j = 2$ for every vertex $j \in P_2$, one gets an instance of TSS(degenerate) for which $OPT = 2$, $\Delta = 3$, and yet the treewidth is linear. Hence to prove theorems~\ref{thm:degenerate} and~\ref{surprising_polytime1} we need a different approach. Our proof is based on considering the vertices of the graph in a natural order that is derived from the degeneracy condition. Using this order we obtain both an upper bound on $OPT$ and a lower bound.  In the case of Theorem~\ref{thm:degenerate}, the bounds differ by a multiplicative factor of at most the maximum over all $t_i$. In the case of Theorem~\ref{surprising_polytime1}, perhaps surprisingly, the bounds match.

Our hardness of approximation results are based on reductions that involve the construction of various gadgets. Like some of the previous hardness results, we first prove hardness results for a variant of TSS on directed graphs, and then reduce from the directed case to the undirected case. In our hardness results we want the graphs to have very low degrees, and one of the aspects that makes the construction of the appropriate gadgets easier is to start from instances of TSS in directed graphs in which vertices have extremal thresholds (either~1, or equal to the in-degree).




\subsection{Discussion}

A major open question in this field is whether the TSS problem can be approximated within some factor of $n^\delta$ for a constant $\delta<1$.
Currently no such approximation algorithm is known even for the TSS($d = 5,t=2$) problem (see Theorem \ref{fiveregthm1}). Likewise, no such algorithm is known even for conservative instances $TSS(d_v > \frac{1}{2}t_v)$, if there is no bound on the maximum degree $\Delta$.

An open question more directly related to our work is whether bounded degree conservative in average (CoA) instances of TSS can be approximated within a ratio of $n^\delta$ for some constant $\delta<1$. By our results, this is indeed true for some subclasses of CoA. For example, Theorem~\ref{alg_mainthm_majority1_statement_main} implies that the $TSS(d=4,t=2)$ problem can be approximated within a ratio of
\[ \min\left(\tilde{O}(OPT),\frac{n}{OPT}\right) = \tilde{O}\left(\sqrt{n}\right)\]
Furthermore we have shown that the $TSS(d=4,t=2)$ problem is $APX$-hard (Theorem \ref{tssdeg4thm2_main}).
Narrowing the gap between these upper and lower bounds remains open, for $TSS(d=4,t=2)$ in particular, and for the weak majority threshold TSS problem in general.

\section{Proofs of algorithmic results}
\subsection{Weak majority threshold functions}
Let $G = (V,E)$ be a graph on $n$ vertices and let $t: V(G) \rightarrow \mathbb{N}$ be a (weak) majority threshold function
satisfying $t(v) \geq d_G(v)/2 $.
We denote the associated target set selection problem by TSS($t_v \geq \frac{1}{2}d_v$).
We build some machinery towards the proof of Theorem \ref{alg_mainthm_majority1_statement_main}.


\BD\label{def:cutwidth} The {\em cutwidth} of graph $G$, denoted by $CW(G)$, is the minimum possible width of a linear ordering of the
vertices of $G$, where the width of an ordering $\sigma$ is the maximum, among all the prefixes of $\sigma$, of
the number of edges that have exactly one vertex in a prefix.
\ED

\BL\label{alg_lemma1} Given a graph $G = (V,E)$ on $n$ vertices and a majority threshold function $t$. If $G$ contains a target set of size $r$ then
$CW(G) \leq \Delta r$ where $\Delta$ is the maximum degree of graph $G$.
\EL
\BPF
Let $T$ be a target set of size $r$ in $G$. Target set $T$ eventually activates all the vertices of $G$. Suppose that the activation order is $v_1,v_2, \ldots, v_n$ where $v_1,v_2, \ldots, v_r$ are the vertices of the target set $T$. As usual we denote by $d_i$ the degree of vertex $v_i$ in $G$ and by $t_i$ the threshold of vertex $v_i$.
For any $1 \leq k \leq n$, let $A_k$ be the set of vertices
 $ \lbrace v_1,v_2, \ldots, v_k \rbrace$ and $B_k$ be the set of vertices $\lbrace v_{k+1},v_{k+2}, \ldots, v_n \rbrace$.
 Denote by $C(k)$ the number of edges in $G$ between the vertices of $A_k$ and the vertices of $B_k$.
 We claim that for each $1 \leq k \leq n$ we have $C(k) \leq \Delta r$.
We will prove this by induction on $k$. This of course holds trivially if $k \leq r$ as the number of edges incident to vertices of $A_k$ is at most $k \Delta$.
Now assume that the claim holds for $k$ and we will prove that the claim holds for $k+1$. By the induction hypothesis the number of edges between $A_k$ and  $B_k$ is at most
$\Delta r$. As vertex $v_{k+1}$ is getting activated in step $k+1$ we know that vertex $v_{k+1}$ has at least $t_{k+1}$ neighbors in $A_k$ and thus
vertex $v_{k+1}$ has at most $d_{k+1} - t_{k+1}$ neighbors in $B_{k+1}$.
We conclude that
\begin{align}\label{majority_main_claim_cutwidth1}
C(k+1) &\leq C(k) + (d_{k+1} - t_{k+1}) - t_{k+1} \notag \\ &= C(k) + d_{k+1} - 2t_{k+1} \notag \\
&\leq C(k) &\text{As $t_{k+1} \geq \frac{1}{2} d_{k+1}$ } \notag
\end{align}
The last inequality follows from the fact that $t$ is a majority threshold function.
\EPF
In \cite{bodlaender1986classes} it is proven that the treewidth of graph $G$ (denoted by $TW(G)$) satisfies \[TW(G) \leq CW(G)\]
For a definition of treewidth (and tree decomposition) see also \cite{bodlaender1986classes}.
We will use the following result from \cite{DBLP:journals/siamcomp/FeigeHL08}.
\BTHM\label{uri_treewidth_result_sqrt}
In any graph of treewidth $w$, a tree decomposition of width at most $O(w \sqrt{\log w})$ can be found in polynomial time in the size of the graph.
\ETHM


Recalling Theorem~\ref{treewidth_theorem_important_related_works}, we now prove Theorem \ref{alg_mainthm_majority1_statement_main}.

\BPF
By Lemma~\ref{alg_lemma1} we have $CW(G) \leq \Delta \cdot OPT$ and hence $TW(G) \leq CW(G) \leq \Delta \cdot OPT$. Thus by Theorem \ref{uri_treewidth_result_sqrt} we can
find in polynomial time a tree decomposition for $G$ of width at most $O(\Delta \cdot OPT \cdot \sqrt{\log (\Delta \cdot OPT)}) )$. Hence by Theorem
 \ref{treewidth_theorem_important_related_works} we can find in polynomial time a target set in $G$ of size at most $O(\Delta \cdot OPT^2 \cdot \sqrt{\log (\Delta \cdot OPT)}) )$.
\EPF

The proof of Theorem~\ref{alg_mainthm_majority1_statement_main}, but with the use of Theorem~\ref{treewidth_theorem_important_related_works} replaced by Theorem~\ref{FPT_result_1}, also proves  Theorem~\ref{alg_mainthm_majority2_statement_main}.



\subsection{Balanced threshold functions}

Recall that a threshold function is balanced if every edge $(u,v)\in E$ satisfies  $\frac{t(u)}{d(u)} + \frac{t(v)}{d(v)} \geq 1$.
Recall also Definition~\ref{def:cutwidth} for the cutwidth of a graph $G$, denoted by $CW(G)$.

Let $M(G) = \max\limits_{v \in G} \frac{d(v)}{t(v)}$. Notice that $M(G) \leq \Delta(G)$.

\BL\label{alg_lemma2} Given a graph $G = (V,E)$ on $n$ vertices and a balanced threshold function $t$, such that $M(G)\geq2$.
If $G$ contains a target set of size $r$ then
$CW(G) \leq (M(G)-1) \Delta r$ where $\Delta$ is the maximum degree of graph $G$.
\EL
\BPF
The case $M=2$ is Lemma \ref{alg_lemma1}, hence we may assume that $M>2$.
Let $T$ be a target set of size $r$ in $G$. Target set $T$ eventually activates all the vertices of $G$. Set $M=M(G)$.
Suppose that the activation order is $v_1,v_2, \ldots, v_n$ where $v_1,v_2, \ldots, v_r$ are the vertices of the target set $T$. As usual we denote by $d_i$ the degree of vertex $v_i$ in $G$ and by $t_i$ the threshold of vertex $v_i$.
For any $1 \leq k \leq n$, let $A_k$ be the set of vertices
 $ \lbrace v_1,v_2, \ldots, v_k \rbrace$ and $B_k$ be the set of vertices $\lbrace v_{k+1},v_{k+2}, \ldots, v_n \rbrace$.
 Denote by $C(k)$ the number of edges in $G$ between the vertices of $A_k$ and the vertices of $B_k$.
 We claim that for each $1 \leq k \leq n$ we have $C(k) \leq (M-1)\Delta r$.
 This of course holds trivially if $k \leq r$ as the number of edges incident to vertices of $A_k$ for $k \leq r$ is at most $\Delta k \leq (M-1)\Delta r$ (as $M>2$).
 Furthermore by the same argument as in Lemma \ref{alg_lemma1} we have for all $k>r$ that
 \[
 C(k) \leq  C(k-1) + d_{k} - 2t_{k}
 \]
 And thus for $k>r$ we have
 \begin{align}\label{what_we_want_to_prove1}
 C(k) &\leq C(r) + \sum_{i=r+1}^{k} (d_i - 2t_i) \notag \\ &\leq \Delta r  + \sum_{i=r+1}^{k} (d_i - 2t_i) &&\text{as $ C(r) \leq \Delta r$ }
 \end{align}
Fix some $k>r$, by (\ref{what_we_want_to_prove1}) all that is left is to prove the following statement.
\begin{equation}\label{what_we_want_to_prove2}
\sum_{i=r+1}^{k} (d_i - 2t_i)  \leq (M-2)\Delta r
\end{equation}
 Let $A \subseteq \{r+1,r+2,\ldots, k \}$ be a set such that for each $i \in A$ we have $\frac{t_i}{d_i} \geq \frac{1}{2}$.
 Let $B =  \{r+1,r+2,\ldots, k \} \setminus A$, notice that for each $i \in B$ we have $\frac{t_i}{d_i} < \frac{1}{2}$.
 As $t$ is a balanced threshold function, the vertices corresponding to set $B$ induce an independent set in graph $G$. \\
 By (\ref{what_we_want_to_prove2}) it is sufficient for us to prove the following.
 \begin{equation}\notag
 \sum_{j \in B} (d_j - 2t_j) - \sum_{i \in A} (2t_i - d_i) \leq  (M-2)\Delta r
\end{equation}
 That is we need to prove that
 \begin{equation}\label{what_we_want_to_prove3}
 \sum_{j \in B} (d_j - 2t_j) \leq  \sum_{i \in A} (2t_i - d_i)  + (M-2)\Delta r
\end{equation}
 As function $t$ is a balanced threshold function we have for all $i \in A$ and $j \in B$ such that $(v_i,v_j)$ is an edge in $G$ that
 $\frac{t_i}{d_i} + \frac{t_j}{d_j} \geq 1$ and by rearranging this inequality we conclude that for all $i \in A$ and $j \in B$ such that $(v_i,v_j)$ is an edge in $G$
 we have
 \begin{equation}\label{balanced_condition1}
   \frac{2t_i - d_i}{d_i-t_i} \geq \frac{d_j-2t_j}{t_j}
 \end{equation}
 Now assign a potential of $M-2$ for every edge touching a vertex in the target set (that is, for each $v_i$ such that $i\in \{1,\ldots,r\}$) and notice
 that the total potential assigned is at most $(M-2)\Delta r$.
 Furthermore for each vertex $v_i$ with $i\in A$ assign a potential of  $\frac{2t_i - d_i}{d_i-t_i}$ for each of the (at most) $d_i-t_i$ edges touching $v_i$ which participate
 in the activation of vertices with indices in $B$ (notice that there are at most $d_i-t_i$ such edges as $t_i$ edges participated in the activation of $v_i$ itself).
 The total potential assigned to edges touching vertices corresponding to indices in $A$ is at most $\sum_{i \in A} (2t_i - d_i)$. Thus the total potential
 assigned to edges touching vertices corresponding to indices in $A$  and edges touching vertices in the target set is
 \[
 \sum_{i \in A} (2t_i - d_i)  + (M-2)\Delta r
 \]
 which is the right hand side of (\ref{what_we_want_to_prove3}).
 Now every vertex $v_j$ such that $j \in B$ has $t_j$ edges which participate in its activation and by the argument above each such edge
 $(v_i,v_j)$ gets a potential of at least
\[
\min \left\{ \frac{2t_i - d_i}{d_i-t_i} , M-2 \right\}
\]
Now notice that  $M-2 \geq  \frac{d_j-2t_j}{t_j}$ by the definition of $M$ and furthermore $\frac{2t_i - d_i}{d_i-t_i} \geq \frac{d_j-2t_j}{t_j}$ by Inequality (\ref{balanced_condition1}). and we conclude that
 \[
\min \left\{ \frac{2t_i - d_i}{d_i-t_i} , M-2 \right\} \geq  \frac{d_j-2t_j}{t_j}
\]
Hence the total potential of edges touching vertex $v_j$ is at least $d_j-2t_j$. This concludes the proof of Inequality (\ref{what_we_want_to_prove3}) and we are done.
\EPF
Using Lemma~\ref{alg_lemma2} instead of Lemma~\ref{alg_lemma1}, the proof of Theorem \ref{alg_mainthm_balanced1_statement_main} is the same as that of Theorem~\ref{alg_mainthm_majority1_statement_main}, and the proof of Theorem \ref{alg_mainthm_balanced2_statement_main} is the same as that of Theorem~\ref{alg_mainthm_majority2_statement_main}.

\subsection{Degenerate threshold functions}\label{degeberate_section1}

Recall that a threshold function
$t: V(G) \rightarrow \mathbb{N}$ is called {\em degenerate} if in every induced subgraph $G'$ of graph $G$ there is a vertex $v$ in $G'$ such that
$t(v) \geq d_{G'}(v)$, where
$d_{G'}(v)$ is the degree of $v$ in $G'$, and $t(v)$ is the threshold of $v$ in graph $G$. Such a function is called a {\em degenerate} threshold  function.
In this section we prove Theorem~\ref{thm:degenerate} concerning the approximability of TSS(degenerate).

For any graph $G$ on $n$ vertices with a degenerate threshold function $t$ there is a degeneracy ordering of its vertices
$(v_1,v_2,\ldots,v_n)$ such that for each $1 \leq i \leq n$ we have $t(v_i) \geq d_p(v_i)$ where $t(v_i)$ is the threshold of vertex $v_i$ and
$d_p(v_i)$ is the number of neighbors of vertex $v_i$ in the vertex set $\{v_1,v_2,\ldots,v_{i-1}\}$ (we define $d_p(v_1)=0$).
Now we present our approximation algorithm for the  TSS(degenerate) problem.
\\ \text{}
\fbox{
\begin{minipage}{14.5 cm}
 \textbf{Algorithm I} \\ \text{} \\
 \textbf{Input:} \hspace{3.5pt} A graph $G = (V,E)$ on $n$ vertices and a degenerate threshold function $t$.\\
 \textbf{Output:} A target set $T$ of size at most $t_{\max} \cdot OPT$ \\ (where $t_{\max}$ is the maximal threshold of function $t$ and $OPT$ is the size of an
 optimal target set).
 \begin{enumerate}
 \item Set $T$ to be an empty set.
 \item Let $(v_1,v_2,\ldots,v_n)$ be a degeneracy ordering of graph $G$.
 \item For each $1 \leq i \leq n$, if $t(v_i) > d_p(v_i)$ then set $T = T \cup \{ v_i \}$.
 \item Return $T$.
 \end{enumerate}
\end{minipage}
}
\\ \text{} \\ \text{} \\
First of all notice that the set $T$ returned by the algorithm above is indeed a target set which activates the whole graph $G$.
This follows from the fact that for $1 \leq i \leq n$, if vertex $v_i$ was not selected by the algorithm then by the property of the degeneracy ordering
we had  $t(v_i) = d_p(v_i)$. Hence after the selection of set $T$ as a target set, the remaining vertices of the graph will be activated in the exact same order as the degeneracy ordering \\
Now we assume that $|T|=r$ and we will prove that $r \leq OPT \cdot t_{\max}$.  \\
By Step $3$ of the algorithm We have that
\begin{align}
  \sum_{i=1}^{n} t(v_i) &\geq \sum_{v \in T} (d_p(v)+ 1 ) + \sum_{v \in V(G) \setminus T} d_p(v) \notag \\
  &\geq r + \sum_{i=1}^{n} d_p(v_i) \notag \\
  &= |E(G)| + r &&\text{(as  $\sum_{i=1}^{n} d_p(v_i) = |E(G)|$ )} \notag
\end{align}
That is
\begin{equation}\label{degenerate_eq1_important}
  \sum_{v \in V(G)} t(v) \geq |E(G)| + r
\end{equation}
Furthermore if $S$ is an optimal target set of size $OPT$ we have that
\begin{equation}\label{degenerate_eq2_important}
  \sum_{v \in V(G) \setminus S} t(v) \leq |E(G)|
\end{equation}
We note that Inequality (\ref{degenerate_eq2_important}) was first observed in \cite{DBLP:journals/tcs/AckermanBW10}.
By Inequality (\ref{degenerate_eq1_important}) we have
\begin{equation}\label{degenerate_eq3_important}
 \sum_{v \in V(G) \setminus S } t(v) + \sum_{v \in S } t(v) \geq |E(G)| + r
\end{equation}
And we conclude from Inequalities (\ref{degenerate_eq3_important}) and (\ref{degenerate_eq2_important}) that
\begin{equation}\label{degenerate_eq4_important} \notag
  \sum_{v \in S } t(v) \geq r
\end{equation}
Hence
$r \leq |S| \cdot t_{\max} = OPT \cdot t_{\max}$ and thus Theorem~\ref{thm:degenerate} is proved.

\section{Hardness of approximation}
\subsection{Hardness of approximation in terms of the maximum degree}
We start by defining the directed target set selection problem, as it is easier to show hardness of approximation for directed variants of the target set selection and then to reduce from those models to the undirected target set selection problem.
The directed target set selection problem is the following:
let $G = (V,E)$ be a directed graph with a threshold function $t: V(G) \rightarrow \mathbb{N}$. Denote by $d^{-}(v)$ the indegree of vertex $v$ in graph $G$ and by
$d^{+}(v)$ the outdegree of vertex $v$.
 We consider the following repetitive process on $G$:
Initially, the states of all vertices are inactive. We pick a
subset $S$ of vertices and activate the vertices of $S$. After that, in each discrete time step, the states of vertices are updated according to following rule: An
inactive vertex $v$ becomes active if at least $t(v)$ (out of its $d^{-}(v)$) in-neighbors become active. The process runs until either all vertices are active or no additional vertices can update states from inactive to active. We call subset $S$ a target set if all the vertices in $G$ have been activated in the end of the process.
We denote this problem as TSS$_D$ .

Let $G = (V,E)$ be a directed graph with a threshold function $t: V(G) \rightarrow \mathbb{N}$. The threshold function $t$ is called an extremal threshold function if for each vertex
 $v \in V(G)$ we have $t(v) = d^{-}(v)$ or $t(v) = 1$. \\
We start by showing a hardness of approximation result for the directed target set selection problem with an extremal threshold function which we denote by TSS$_D$($t_v \in \{1,d_v \}$).
In Theorem $3.1$ of \cite{DBLP:journals/computability/BazganCNS14} the following is proven (it is not stated explicitly there that the threshold function is extermal but it follows from their reduction).
\BL\label{directedTSSlem1}
 The  TSS$_D$($t_v \in \{1,d_v \}$) problem cannot be approximated within the ratio $f(OPT)$ for any computable function $f$, unless $FTP = W[P]$.
\EL
The reduction in the Theorem $3.1$ of \cite{DBLP:journals/computability/BazganCNS14} is from the Monotone Circuit Satisfiability  problem and it preserves the exact value of the solution.
In \cite{DBLP:journals/ipl/DinurS04} it is shown that the Monotone Circuit Satisfiability problem is $NP$-hard to approximate within the ratio $O\left(2^{\log^{1-\epsilon} n}\right)$
for any fixed constant $\epsilon>0$. Hence we have the following.
\BL\label{directedTSSlem11}
 The  TSS$_D$($t_v \in \{1,d_v \}$) problem is $NP$-hard to approximate within the ratio $O\left(2^{\log^{1-\epsilon} n}\right)$
for any fixed constant $\epsilon>0$.
\EL
Given a directed graph we denote by $d^t(v)$ the total degree of vertex $v$, that is $d^t(v) = d^-(v) + d^+(v)$.
\BL\label{directedTSSlem2}
 The TSS$_D$($d^t \leq 3$,$t \in \{1,2\}$) problem
 cannot be approximated within the ratio $O\left(2^{\log^{1-\epsilon} n}\right)$
for any fixed constant $\epsilon>0$, unless $P=NP$.
\EL
\BPF
We will reduce from the TSS$_D$($t_v \in \{1,d_v \}$) problem.
Let $(G,t)$ be an instance of the  TSS$_D$($t_v \in \{1,d_v \}$) problem.
We will assume without loss of generality that for any vertex $v \in V(G)$ we have $d^+(v) \geq 1$ and $d^-(v) \geq 1$ .
\\
If graph $G$ contains  a vertex $v$ such that $d^t(v) > 3$ we will replace vertex $v$ with the following gadget: \\
Let $u_1,u_2,\ldots,u_k$ be the in-neighbors of vertex $v$ in $G$ (where $k=d^-(v)$).
Let $w_1,w_2,\ldots,w_l$ be the out-neighbors of vertex $v$ in $G$ (where $l=d^+(v)$). \\
Now we remove vertex $v$ from graph $G$ and add vertices $u'_1,u'_2,\ldots,u'_k$ and $w'_1,w'_2,\ldots,w'_l$ to graph $G$.
Now we add directed edges to graph $G$ in the following manner.
\begin{itemize}
  \item For all $1 \leq i \leq k-1$ we add a directed edge $(u'_i,u'_{i+1})$ (that is $u'_i$ is an in-neighbor of $u'_{i+1}$).
  \item For all $1 \leq i \leq l-1$ we add a directed edge $(w'_i,w'_{i+1})$.
  \item We add a directed edge $(u'_k,w'_1)$.
  \item For all $1 \leq i \leq k$ we add a directed edge $(u_i,u'_i)$.
  \item For all $1 \leq i \leq l$ we add a directed edge $(w'_i,w_i)$.
\end{itemize}
Notice that the induced subgraph of $G$ consisting of vertices $u'_1,u'_2,\ldots,u'_k$ and $w'_1,w'_2,\ldots,w'_l$ is a directed path.
Now we add thresholds for the vertices $u'_1,u'_2,\ldots,u'_k$ and $w'_1,w'_2,\ldots,w'_l$.
If $t(v) = 1$ then we set the thresholds in the following manner.
\begin{itemize}
  \item For all $1 \leq i \leq k$ set $t(u'_i)=1$.
  \item For all $1 \leq i \leq l$ set $t(w'_i)=1$.
\end{itemize}
If $t(v) = d^-(v)$ then we set the thresholds in the following manner.
\begin{itemize}
  \item Set $t(u'_1)=1$.
  \item For all $2 \leq i \leq k$ set $t(u'_i)=2$.
  \item For all $1 \leq i \leq l$ set $t(w'_i)=1$.
\end{itemize}
Denote the resulting graph by $G'$ and we will denote by $G$ the original graph before the replacement operation.
If $G$ has a target set $T$ such that $v \not \in T$ then $T$ is also a target set of $G'$. On the other hand if  $G$ has a target set $T$  such that $v \in T$ then we remove vertex $v$
 from set $T$. Now notice that $T \bigcup \{w'_1\}$ is a target set of $G'$. We conclude that if $G$ has a target set of size $q$ then $G'$ has a corresponding target set of size $q$. \\
Now assume that $G'$ has a target set $T'$. If for all $1 \leq i \leq k$ we have $u'_i \not \in T'$ and for all $1 \leq i \leq l$ we have $w'_i \not \in T'$ then $T'$ is also a target
 set of $G$. Otherwise we do the following
\begin{itemize}
  \item For each $1 \leq i \leq k$ if $u'_i \in T'$ then let $T' = T' \setminus \{u'_i\}$.
  \item For each $1 \leq i \leq l$ if $w'_i \in T'$ then let $T' = T' \setminus \{w'_i\}$.
\end{itemize}
Now notice that the set $T' \bigcup \{v\}$ is a target set of $G$.
Thus if $G'$ has a target set of size $q$ then $G$ has a corresponding target set of size at most $q$.
We conclude that $G$ has a target set of size at most $q$ if and only if $G'$ has a target set of size at most $q$. \\
Now if graph $G'$ contains a vertex $v'$ such that $d^t(v') > 3$ we will replace vertex $v'$ with the same gadget as described above. We do this replacement operation iteratively until we finally get a graph $H$ such that each vertex $v \in H$ satisfies $d^t(v) \leq 3$. The theorem follows from the following observations.
\begin{enumerate}
  \item Each threshold of a vertex in graph $H$ is at most $2$ (as the in-degree and out-degree of each vertex in $H$ is at most $2$).
  \item Graph $G$ has a target set of size at most $q$ if and only if graph $H$ has a target set of size at most $q$.
  \item Given a target set of size $q$ in graph $H$ one can convert it (in polynomial time) to a target set of size at most $q$ in graph $G$.
\end{enumerate}
And thus we are done.
\EPF
\BL\label{undirectedTSSlem11}
The TSS($d \leq 3$,$t \in \{1,2\}$) problem cannot be approximated within the ratio $O\left(2^{\log^{1-\epsilon} n}\right)$
for any fixed constant $\epsilon>0$, unless $P=NP$.
\EL
\BPF
We will reduce from the TSS$_D$($d^t \leq 3$,$t \in \{1,2\}$) problem (the problem addressed in Lemma \ref{directedTSSlem2}).
Let $(G,t)$ be an instance of the TSS$_D$($d^t \leq 3$,$t \in \{1,2\}$) problem.
We replace each directed edge $e$ of graph $G$ by a gadget $H$ consisting of $4$ vertices and $7$ undirected edges.
Assume that $e=(v_1,v_2)$. Remove edge $e$ from graph $G$ and add vertices $u_1,u_2,u_3,u_4$ with the undirected edges:
\[
(u_1,v_1) , (u_1,u_2) , (u_1,u_3) , (u_2,u_3) , (u_4,u_2) , (u_4,u_3) , (u_4,v_2)
\]
Now set the following thresholds
\[
t(u_1) = 1 , t(u_2) = 1 , t(u_3) = 2 , t(u_4) = 2
\]
We stress that for each edge $e$ of graph $G$ we add four different vertices (hence if graph $G$ contains $|E|$ directed edges we add $4|E|$ vertices).


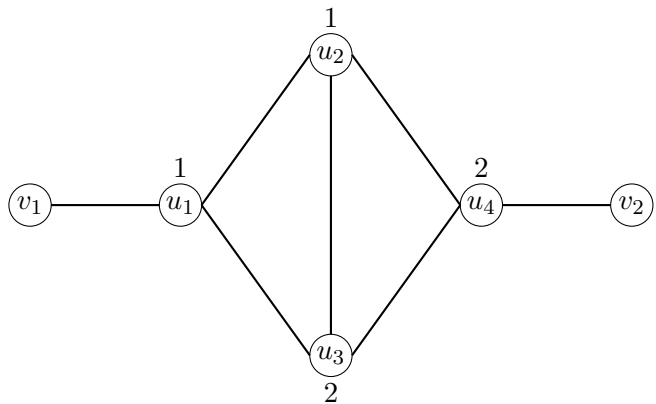
\begin{figure}
  \centering

\begin{tikzpicture}
\draw (1,0) circle (8pt);
\draw (3,0) circle (8pt);
\draw (5,2) circle (8pt);
\draw (5,-2) circle (8pt);
\draw (7,0) circle (8pt);
\draw (9,0) circle (8pt);

\node at (1,0) {$v_1$};
\node at (3,0) {$u_1$};
\node at (3,0.5) {$1$};
\node at (5,2) {$u_2$};
\node at (5,2.5) {$1$};
\node at (5,-2) {$u_3$};
\node at (5,-2.5) {$2$};
\node at (7,0) {$u_4$};
\node at (7,0.5) {$2$};
\node at (9,0) {$v_2$};


\draw[thick] (1.28,0) -- (2.72,0);
\draw[thick] (7.28,0) -- (8.72,0);
\draw[thick] (3.28,0) -- (4.72,2);
\draw[thick] (3.28,0) -- (4.72,-2);
\draw[thick] (5.28,2) -- (6.72,0);
\draw[thick] (5.28,-2) -- (6.72,0);
\draw[thick] (5,-1.72) -- (5,1.72);
\end{tikzpicture}
 \caption{Gadget $H$}\label{figure_111}
\end{figure}


After replacing edge $e=(v_1,v_2)$ by the gadget $H$ above we notice that if $v_1$ gets activated than the gadget $H$ will activate vertex $v_2$. On the other hand if vertex $v_2$ is activated and none of the vertices in gadget $H$ are active then vertex $v_1$ will not get activated by the gadget $H$. Furthermore if a vertex in gadget $H$ belong to a target set we can replace it with vertex $v_1$. We note that an almost identical gadget appears in the paper
\cite{DBLP:journals/computability/BazganCNS14}.
\EPF
\BTHM\label{cubicthmtargetset1}
(Restatement of Theorem \ref{cubicthmtargetset1_main})
The TSS($d = 3$,$t \in \{1,2\}$). problem cannot be approximated within the ratio $O\left(2^{\log^{1-\epsilon} n}\right)$
for any fixed constant $\epsilon>0$, unless $P=NP$.
\ETHM
\BPF
We will reduce from the TSS($d \leq 3$,$t \in \{1,2\}$) problem (the problem addressed in Lemma \ref{undirectedTSSlem11}).
Let $(G,t)$ be an instance of the TSS($d \leq 3$,$t \in \{1,2\}$) problem,
where $G$ is an undirected graph on $n$ vertices $v_1,v_2,\ldots,v_n$.
 Assume without loss of generality that $G$ has no vertices of degree $1$. We shall take two disjoint copies $H_1,H_2$ of graph $G$,. The vertices of $H_1$ are $u_1,u_2,\ldots,u_n$ and the vertices of $H_2$ are $w_1,w_2,\ldots,w_n$ where for each $1 \leq i \leq n$ vertex $u_i$ corresponds to vertex $v_i$ and vertex $w_i$ corresponds to vertex $v_i$. Graph $G$ has an edge $(v_i,v_j)$ if and only if $H_1$ has an edge $(u_i,u_j)$ and $H_2$ has an edge $(w_i,w_j)$.
Let $G' = H_1 \bigcup H_2$, that is $G'$ is the disjoint union of graphs $H_1$ and $H_2$.
For all $1 \leq i \leq n$ if we have $d(v_i)=2$ in graph $G$ then we add an edge $(u_i,w_i)$ to graph $G'$. Graph $G'$ is cubic after this operation.
Now notice that the following holds.
\begin{enumerate}
  \item If $S \subseteq [1,n]$ is a set of indices such that $\{v_i | i \in S \}$ is a target set of $G$ then $\{u_i | i \in S \} \bigcup \{w_i | i \in S \}$ is a target set of $G'$.
  \item If $S_1 \subseteq [1,n]$ and $S_2 \subseteq [1,n]$ are sets of indices such that $\{u_i | i \in S_1 \}  \bigcup \{w_i | i \in S_2 \} $ is a target set of $G'$ then
  $\{v_i | i \in S_1 \bigcup S_2 \}$ is a target set of $G$.
\end{enumerate}
We have shown that every target set of size $k$ in $G$ corresponds to a target set of size $2k$ in $G'$ and that every target set of size $k$ in $G'$ corresponds to a target set of size at most $k$ in $G$.
\EPF
We note that by the same chain of reductions we can prove Theorem \ref{cubicthmtargetset1OPT}. 
\BTHM\label{fiveregthm1}
The TSS($d = 5$,$t = 2$) problem 
cannot be approximated within the ratio $O\left(2^{\log^{1-\epsilon} n}\right)$
for any fixed constant $\epsilon>0$, unless $P=NP$.
\ETHM
\BPF
(Sketch) We reduce from the TSS($d = 3$,$t \in \{1,2\}$) problem (the problem addressed in Theorem \ref{cubicthmtargetset1}).
Let $G$ be an undirected cubic graph with thresholds at most $2$ in which we want to find a target set. We add to graph $G$ a "super" vertex gadget as shown in \cite{chen1}
(see Figure $9$ in  \cite{chen1}), and we connect each vertex of threshold $1$ in the original graph $G$ to this gadget.
 The resulting graph is of degree at most $5$ and thresholds exactly $2$. Now we can apply techniques similar to the ones in Theorem \ref{cubicthmtargetset1} to create a corresponding $5$-regular graph.
\EPF
\BCR\label{degenerate_colorally13}
(Restatement of Theorem \ref{degenerate_colorally13_main}) The TSS(degenerate) problem cannot be approximated within the ratio $O\left(2^{\log^{1-\epsilon} n}\right)$
for any fixed constant $\epsilon>0$, unless $P=NP$.
\ECR
\BPF
We reduce from the TSS($d = 5$,$t = 2$) problem. Let $(G,t)$ be an instance of the TSS($d = 5$,$t = 2$) problem, where $G$ is a graph on $n$ vertices
 $v_1,v_2,\ldots,v_n$.
We create a graph $G'$ by adding to graph $G$ three vertices $w_1,w_2,w_3$ and connecting each such vertex to all the vertices of the original graph $G$.
Furthermore we create a threshold funcion $t'$ which satisfies the following conditions.
\begin{enumerate}
  \item  $t'(w_1)=t'(w_2)=t'(w_3)=n$
  \item  $t'(v_1)=t'(v_2)=\ldots = t'(v_n)=5$
\end{enumerate}
And we are done as $(G',t')$ is an instance of the TSS(degenerate) problem.
\EPF
\BTHM\label{tssdeg4thm2}
(Restatement of Theorem \ref{tssdeg4thm2_main})
The TSS($d = 4$,$t =2$) problem is $APX$-hard.
\ETHM
\BPF
(Sketch) The TSS($d=3$,$t=3$) problem is equivalent to the cubic vertex cover problem which is APX-hard (\cite{DBLP:conf/ciac/AlimontiK97}).
The TSS($d=3$,$t=3$) is equivalent to the  TSS$_D$($d^{-}=3$,$d^{+}=3$,$t=3$) problem (this follows by replacing each edge in the undirected graph by two anti-parallel directed edges).
Now we can reduce from this problem to the
TSS($d = 3$,$t \in \{1,2\}$) problem
in which the size of the target set is linear in the size of the graph (this can be done by reductions similar to those in Lemma
 \ref{directedTSSlem2} and Lemma \ref{undirectedTSSlem11}). \\
 Let $(G,t)$ be the instance of the TSS($d = 3$,$t \in \{1,2\}$) problem in which the size of the target set is linear in the size of the graph. Assume that $G$ contains $n$ vertices
 $v_1,v_2,\ldots,v_n$.
 We shall take two disjoint copies $H_1,H_2$ of graph $G$,. The vertices of $H_1$ are $u_1,u_2,\ldots,u_n$ and the vertices of $H_2$ are $w_1,w_2,\ldots,w_n$ where for each $1 \leq i \leq n$ vertex $u_i$ corresponds to vertex $v_i$ and vertex $w_i$ corresponds to vertex $v_i$. Graph $G$ has an edge $(v_i,v_j)$ if and only if $H_1$ has an edge $(u_i,u_j)$ and $H_2$ has an edge $(w_i,w_j)$.
Let $G' = H_1 \bigcup H_2$, that is $G'$ is the disjoint union of graphs $H_1$ and $H_2$.
For all $1 \leq i \leq n$ if we have $t(v_i)=2$ in graph $G$ then we add an edge $(u_i,w_i)$ to graph $G'$.
We assume without loss of generality that graph $G$ is connected. Now notice that the following holds.
\begin{enumerate}
  \item If $S \subseteq [1,n]$ is a set of indices such that $\{v_i | i \in S \}$ is a target set of $G$ then $\{u_i | i \in S \} \bigcup \{w_1 \}$ is a target set of $G'$.
  \item If $S_1 \subseteq [1,n]$ and $S_2 \subseteq [1,n]$ are sets of indices such that $\{u_i | i \in S_1 \}  \bigcup \{w_i | i \in S_2 \} $ is a target set of $G'$ then
  $\{v_i | i \in S_1 \bigcup S_2 \}$ is a target set of $G$.
\end{enumerate}
We have shown that every target set of size $k$ in $G$ corresponds to a target set of size $k+1$ in $G'$ and that every target set of size $k$ in $G'$ corresponds to a target set of size at most $k$ in $G$. Notice that each vertex of threshold $2$ in $G'$ is of degree $4$ and each vertex of threshold $1$ in $G'$ is of degree $3$.
Let $S_3 \subseteq [1,n]$ be the set of indices such that $t(v_i)=1$ for all $i \in S_3$.
For each $i \in S_3$ we take vertices $u_i,w_i$ in graph $G'$ and connect them to the following ladder gadget which we denote as gadget $L$
(we stress that for each $i \in S_3$ we use a different ladder gadget)


\begin{figure}
\centering

\begin{tikzpicture}
\draw (1,0) circle (8pt);
\draw (1,2) circle (8pt);
\draw (3,0) circle (8pt);
\draw (3,2) circle (8pt);
\draw (5,0) circle (8pt);
\draw (5,2) circle (8pt);
\draw (7,0) circle (8pt);
\draw (7,2) circle (8pt);
\draw (9,0) circle (8pt);
\draw (9,2) circle (8pt);
\draw (3,-2) circle (8pt);
\draw (5,-2) circle (8pt);

\node at (1,0) {$s_0$};
\node at (1,2) {$t_0$};
\node at (3,0) {$s_1$};
\node at (3,2) {$t_1$};
\node at (5,0) {$s_2$};
\node at (5,2) {$t_2$};
\node at (7,0) {$s_3$};
\node at (7,2) {$t_3$};
\node at (9,0) {$s_0$};
\node at (9,2) {$t_0$};
\node at (3,-2) {$u_i$};
\node at (5,-2) {$w_i$};


\draw[thick] (1.28,0) -- (2.72,0);
\draw[thick] (1.28,0) -- (2.72,2);
\draw[thick] (1.28,2) -- (2.72,0);
\draw[thick] (1.28,2) -- (2.72,2);
\draw[thick] (3.28,2) -- (4.72,2);
\draw[thick] (3.28,0) -- (4.72,2);
\draw[thick] (3.28,2) -- (4.72,0);
\draw[thick] (5.28,0) -- (6.72,0);
\draw[thick] (5.28,2) -- (6.72,2);
\draw[thick] (5.28,0) -- (6.72,2);
\draw[thick] (5.28,2) -- (6.72,0);
\draw[dashed] (7.28,0) -- (8.72,0);
\draw[dashed] (7.28,2) -- (8.72,2);
\draw[dashed] (7.28,0) -- (8.72,2);
\draw[dashed] (7.28,2) -- (8.72,0);
\draw[thick] (3,-0.28) -- (3,-1.72);
\draw[thick] (5,-0.28) -- (5,-1.72);
\end{tikzpicture}
\caption{Gadget $L$} \label{gif_222}
\end{figure}
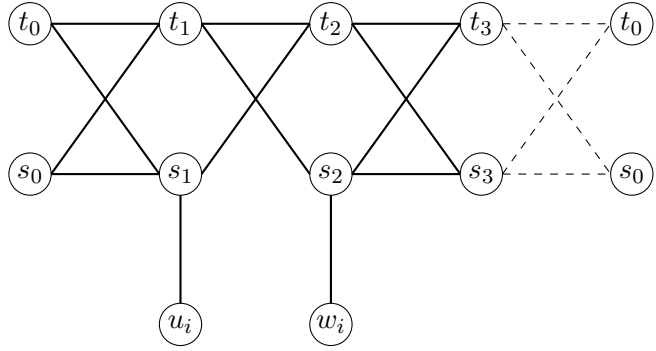

Gadget $L$ consist of vertices $t_0,t_1,t_2,t_3$ and $s_0,s_1,s_2,s_3$ and edges
\[
(t_i,t_{(i+1) \mod 4}) , (s_i,s_{(i+1) \mod 4}) , (s_i,t_{(i+1) \mod 4}) , (t_i,s_{(i+1) \mod 4})
\]
 for all $0 \leq i \leq 3$. Now we remove edge $(s_1,s_2)$ and connect $s_1$ to a vertex $u_i$ and $s_2$ to vertex $w_i$.
Finally we set the threshold of vertices $s_0,s_1,s_2,s_3,t_0,t_1,t_2,t_3,u_i,w_i$ in $G'$ to be $2$.
Now notice that it is sufficient to activate two vertices in gadget $L$ in order to activate all its vertices, and that it is necessary to activate at least two vertices in gadget $L$ even if vertices $u_i$ and $w_i$ are activated in order to activate all its vertices.
The resulting graph $G'$ is $4$-regular with all vertices of threshold $2$ and thus we are done.
\EPF
\BTHM\label{feedbackvertexset_4regular}
(Restatement of Theorem \ref{feedbackvertexset_4regular_main})
The TSS($d=4$,$t=3$) problem is $APX$-hard (that is feedback vertex set is $APX$-hard on $4$-regular graphs).
\ETHM
\BPF
In \cite{DBLP:journals/algorithmica/Rizzi09} it is shown that the TSS($t_v = d_v-1$,$d \leq 4$) problem (that is feedback vertex set in graphs of maximum degree $4$)
is $APX$-hard even if the size of the target set is linear in the size of the graph. Let graph $(G,t)$ be an instance of
 the TSS($t_v = d_v-1$,$d \leq 4$) problem (in which the size of the target set is linear in the size of the graph), where $G$ is a graph on $n$ vertices $v_1,v_2,\ldots,v_n$.
 We create a graph $G'$ by taking two disjoint copies $H_1,H_2$ of graph $G$. The vertices of $H_1$ are $u_1,u_2,\ldots,u_n$ and the vertices of $H_2$ are $w_1,w_2,\ldots,w_n$.
 Now as long as there is an index $i$ such that $d(u_i)=d(w_i)<4$ in graph $G'$, we do the following: we add a gadget $L$ as described in the proof of Theorem \ref{tssdeg4thm2} to graph $G'$
 and connect this gadget to vertices $u_i$ and $w_i$. Furthermore we set the threshold of all the vertices in the gadget $L$ to be $3$ and we set $t(u_i)=t(u_i)+1$ and
 $t(w_i)=t(w_i)+1$. Now notice that it is sufficient to activate $3$ vertices in gadget $L$ in order to activate all its vertices, and that it is necessary to activate at least $3$ vertices in gadget $L$ even if vertices $u_i$ and $w_i$ are activated in order to activate all its vertices.
 The resulting graph $G'$ is $4$-regular and all its thresholds are $3$ and thus we are done.
\EPF

\subsection{Hardness of approximation in terms of the average degree}
Once again we start from a variant of the directed target set selection problem, as it is easier to show hardness of approximation for directed variants of the target set selection and then to reduce from those models to the undirected target set selection problem.
We will need the following definitions. The directed feedback vertex set (DFVS) problem is the following problem:
given a directed graph $G = (V,E)$ we wish to delete the minimum number of vertices in $G$ so that the resulting graph is acyclic (that is without directed cycles).
We shall denote a set of vertices whose removal makes directed graph $G$ acyclic a feedback vertex set.
Let $G = (V,E)$ be a directed graph with a threshold function $t: V(G) \rightarrow \mathbb{N}$. The threshold function $t$ is called a unanimous threshold function if for each vertex
 $v \in V(G)$ we have $t(v) = d^{-}(v)$.
\BTHM\label{fvsistss1}
Let $(G,t)$ be an instance of the TSS$_D$($t_v = d_v^{-}$) problem. Then
a vertex set $S \subseteq V(G)$ is a feedback vertex set of the directed graph $G$ if and only if the set $S$ is a target set of the instance $(G,t)$.
\ETHM
\BPF
Assume that directed graph $G=(V,E)$ has $n$ vertices. Let $(G,t)$ be an instance of the TSS$_D$(${t_v = d_v^{-}}$) problem.
If a vertex set $S \subseteq V$ is a target set for the instance $(G,t)$
then the induced subgraph $G'$ of $G$ on vertices $V \setminus S$ is acyclic for if we assume by contradiction that $G'$ contains a cycle $C$ then it follows from the fact that $t$ is a unanimous threshold function that the vertices of cycle $C$ will not be activated by the target set $S$. We conclude that $S$ is a feedback vertex set of directed graph $G$.

Now assume that the vertex set $T \subseteq V$ is a feedback vertex of size $k$ of directed graph $G$. We can topologically sort the vertices of the induced subgraph $G'$ of $G$ on vertices $V \setminus S$, thus getting an ordering $v_1,v_2,\ldots,v_n$  of the vertices of $G$ where the following holds.
\begin{enumerate}
  \item the vertices $v_1,\ldots,v_k$ are the vertices of $T$.
  \item For all $k<i\leq n$ the following holds: if $v_j$ is an in-neighbor of $v_i$ then $j<i$ (that is all the in-neighbors of $v_i$ appear before it in the ordering).
\end{enumerate}
We denote this ordering by $O$.
Now we notice that if in the instance $(G,t)$ of the TSS$_D$(${t_v = d_v^{-}}$) problem
vertices $v_1,\ldots,v_k$ in ordering $O$ are activated then vertices $v_{k+1},v_{k+2},\ldots,v_n$ can be activated in this order and thus $T$ is a target set of the instance $(G,t)$.
\EPF
In \cite{DBLP:journals/combinatorica/Seymour95} it is shown that
the DFVS problem can be approximated within a ratio of $\tilde{O}(\log |V|)$.
The hardness of approximation result in \cite{DBLP:journals/siamcomp/GuruswamiHMRC11} implies that it is $NP$-hard to approximate the DFVS problem within any
constant factor assuming the $UGC$ (this result is also proven in \cite{DBLP:journals/toc/Svensson13} and \cite{DBLP:journals/toc/GuruswamiL16}).
We call a directed graph $G$ restricted if for each vertex $v \in V(G)$ one of the following two conditions holds.
\begin{enumerate}
  \item $d^-(v)=1$ and $d^+(v)=2$.
  \item $d^+(v)=1$ and $d^-(v)=2$.
\end{enumerate}
and furthermore the number of vertices of indegree $1$ in $G$ equals the number of vertices of outdegree $1$ in $G$.
We shall require the following theorem which we prove in appendix \ref{appendix1DFVS}.
\BTHM\label{restrictedgraphtheorem0}
 It is $NP$-hard to approximate the DFVS problem on restricted directed graphs within any
constant factor assuming the $UGC$.
\ETHM
\BCR
The target set selection problem on undirected cubic graphs in which half the vertices have threshold $1$ and the other half threshold $2$ is $NP$-hard to approximate within any
constant factor assuming the $UGC$.
\ECR
\BPF
 We reduce from the DFVS problem on restricted directed graphs. Given a graph $G$ which is an instance of the DFVS problem on restricted graphs we have by Theorem
 \ref{fvsistss1} an instance $(G,t)$ of directed target set selection on restricted graphs with unanimous thresholds.
 We replace each directed edge of $G$ by the gadget introduced in Lemma \ref{undirectedTSSlem11}.
\EPF
\BTHM\label{avgdegthm1}
(Restatement of Theorem \ref{avgdegthm1_main})
The TSS($\bar{d}=4$,$d \leq 5$,$t=2$) problem is $NP$-hard to approximate within any
constant factor assuming the $UGC$.
\ETHM
\BPF
(Sketch) We reduce from the target set selection problem on undirected cubic graphs in which half the vertices have threshold $1$ and the other half threshold $2$.
Let $(G,t)$ be an instance of the target set selection problem on undirected cubic graphs in which half the vertices have threshold $1$ and the other half threshold $2$.
We apply the same gadget on $G$ as in Theorem \ref{fiveregthm1}.
\EPF

\section{Target Set Selection with Partial Incentives}\label{partial_incentives_model_Sect}

Let $G = (V,E)$ be a graph on $n$ vertices $v_1,v_2,\ldots,v_n$ and let $t$ be a threshold function
$t: V(G) \rightarrow \mathbb{N}_0$. We assume that for all $i$ we have $t(v_i) \leq d(v_i)$, where $d(v_i)$ is the degree of the vertex. \\
The following optimization problem is called Target Set Selection with Partial Incentives (\cite{DBLP:conf/sirocco/CordascoGRV15}): Given a graph $G$ and a threshold function $t$, find
a partial incentive function $q: V(G) \rightarrow \mathbb{N}_0$ such that the graph $G$ with the threshold function $p=t-q$ is activated by the target set consisting of vertices
of threshold $0$ in $p$, and we want to find such function $q$ that minimizes $\sum_{i=1}^{n} q(v_i)$, which we denote as the weight of function $q$.
Henceforth we denote this problem by TSS$_P$.
A $k$-approximate solution to an instance of the TSS$_P$ problem returns a partial incentive function of weight at most
$k \cdot OPT$ where $OPT$ is the weight of an optimal solution for this instance..
\BTHM\label{simple_pit1}
Given an instance of $(G,t)$ of TSS$_P$, where $G$ is a graph on $n$ vertices, a partial incentive function $q$ of weight at most $n \sqrt{2 OPT}$ can be found in polynomial time,
 where $OPT$ is the weight of an optimal solution for this instance.
\ETHM
\BPF
In Corollary $1$ of \cite{cordasco2018whom} it is proven that if $t_{\min}$ is the minimum threshold of function $t$ then $OPT \geq \frac{1}{2} t_{\min} ( t_{\min} + 1)$.
Hence we have $t_{\min} \leq \sqrt{2 OPT}$. We activate a vertex with threshold $t_{\min}$ thus paying at most $\sqrt{2 OPT}$ units and call the resulting graph $G'$ and the resulting threshold function $t'$.
Let $t'_{\min}$ be the minimum threshold of function $t'$ and notice that as before $t'_{\min} \leq \sqrt{2 OPT}$ as the optimum can only decrease in the
instance ($G'$,$t'$) resulting by an activation of a vertex in graph $G$. Continuing iteratively we activate the whole graph $G$ by paying at most $\sqrt{2 OPT}$ units for each vertex and thus we are done.
\EPF

Let $t$ be a threshold function
$t: V(G) \rightarrow \mathbb{N}$ such that $t(v) \geq \lceil d_G(v)/2 \rceil $ where
$d_G(v)$ is the degree of $v$ in $G$, recall that  such function is called a majority threshold  function. \\
We denote by  TSS$_P$($t_v \geq \frac{1}{2}d_v$) the Target Set Selection with Partial Incentives under majority thresholds.
We start by presenting a polynomial time algorithm which returns an $\tilde{O}(OPT)$-approximate solution to the  TSS$_P$($t_v \geq \frac{1}{2}d_v$) problem. That is we find in polynomial time a partial incentive function  of weight
$\tilde{O}(OPT^2)$.

Recall definition~\ref{def:cutwidth} regarding the cutwidth of graph $G$, denoted by $CW(G)$.
\BL\label{pit_alg_lemma1} Given a graph $G = (V,E)$ on $n$ vertices, if the instance ($G$,$t$) of
the TSS$_P$($t_v \geq \frac{1}{2}d_v$) problem
has a solution of weight $r$ then
$CW(G) \leq 2 r$.
\EL
\BPF
Let $q$ be an optimal partial incentive function for the instance $(G,t)$ of the TSS$_P$(${t_v \geq \frac{1}{2}d_v}$) problem.
Suppose that the activation order of the vertices under threshold function $p=t-q$ is
$v_1,v_2, \ldots, v_n$.
As usual we denote by $d_i$ the degree of vertex $v_i$ in $G$, $t_i=t(v_i)$ and $q_i=q(v_i)$ and $p_i = t(v_i) - q(v_i)$.
For any $1 \leq k \leq n$, let $A_k$ be the set of vertices
 $ \lbrace v_1,v_2, \ldots, v_k \rbrace$ and $B_k$ be the set of vertices $\lbrace v_{k+1},v_{k+2}, \ldots, v_n \rbrace$.
 Denote by $C(k)$ the number of edges in $G$ between the vertices of $A_k$ and the vertices of $B_k$.
 We claim that for each $1 \leq k \leq n$ we have $C(k) \leq 2 r$.
 Notice that by the same argument as in Lemma \ref{alg_lemma1} we have for all $1 \leq k \leq n$ that
 \[
 C(k) \leq  C(k-1) + d_{k} - 2p_{k}
 \]
 And thus for $1 \leq k \leq n$ we have
 \begin{align}\label{pim_what_we_want_to_prove1}
 C(k) &\leq  \sum_{i=1}^{k} (d_i - 2p_i) \notag \\ &= \sum_{i=1}^{k} (d_i - 2t_i + 2q_i) &&\text{as $ p_i = t_i - q_i$ } \notag \\
 &\leq 2\sum_{i=1}^{k} q_i &&\text{as $t_i \geq \frac{d_i}{2} $ since $t$ is a majority threshold function}  \notag \\
 &\leq 2r  &&\text{as $r = \sum_{i=1}^{n} q_i$ }  \notag \\
 \end{align}
This concludes the proof.
\EPF
Given a graph $G$ on $n$ vertices, a cutwidth arrangement that approximates the optimal cutwidth arrangement within a factor of $O(\log^{3/2} n)$
can be found in polynomial time (see for example \cite{casel2019graph}). That is we can find in polynomial time a cutwidth arrangement $(v_1,v_2,\ldots,v_n)$ of the vertices of $G$ such that for all $1 \leq k \leq n$ the number of
edges between the set vertices $\{ v_1,v_2,\ldots,v_k \}$ and the set of vertices   $\{ v_{k+1},v_2,\ldots,v_n \}$ is at most $O(\log^{3/2} n \cdot CW(G))$, where $CW(G)$ is the cutwidth of $G$.
Now we present our first polynomial time approximation algorithm for the TSS$_P$($t_v \geq \frac{1}{2}d_v$) problem.
As usual we denote $q_i = q(v_i)$ for notational clarity.
\\ \text{} \\
\fbox{
\begin{minipage}{14.5 cm}
 \textbf{Algorithm II} \\ \text{} \\
 \textbf{Input:} \hspace{3.5pt} A graph $G = (V,E)$ on $n$ vertices and a majority threshold function $t$.\\
 \textbf{Output:} A partial incentive function $q$ of weight at most $O( \log^{3/2} n \cdot OPT^2 )$ \\ (where $OPT$ is the weight of an optimal solution).
 \begin{enumerate}
 \item Let $(v_1,v_2,\ldots,v_n)$ be a cutwidth arrangement of graph $G$ of width at most $W = O( \log^{3/2} n \cdot CW(G) )$.
 \item Set $i=0$.
 \item Set $i=i+1$.
 \item If activating the vertex set $\{v_{i+1},v_{i+2},\ldots,v_n\}$ as a target set activates the whole graph then goto step $3$
  (when we say that we activate a vertex we mean it as in the standard model of the target set selection problem).
 \item For each $1\leq j < i$ do the following: set $q_j = \min\{t_j,n_j\}$ where $n_j$ is the number of neighbors vertex $v_j$ has in the vertex set $\{v_{i},v_{i+1},\ldots,v_n\}$ and $t_j = t(v_j)$.
 \item Set $q_i = t_i$.
 \item Let $(G',t')$ be an instance of the TSS$_P$ problem
 where graph $G'$ is the induced subgraph of $G$ on vertices $v_{i+1},v_{i+2},\ldots,v_n$  and
  threshold function $t'$ for the vertices of $G'$ is the threshold function resulting from activating vertices $v_1,v_2,\ldots,v_i$ of the instance $(G,t)$.
\item Renumber the vertices of $G'$ to be $v_1,v_2,\ldots,v_{n-i}$. Set $G=G'$,$t=t'$, $n=n-i$ and goto step 1 if $n > 0$.
 \end{enumerate}
\end{minipage}
}
\\ \text{} \\ \text{} \\

\BTHM\label{approx_of_alg_partial_incentive0}
Algorithm II returns a valid partial incentive function $q$.
\ETHM
\BPF
We will prove that the target set $S$ consisting of vertices of threshold $0$ of the function $p=t-q$ activates all the vertices of graph $G$ (with threshold function $p$).
This follow from the fact that each time after we execute steps $5,6$ of the algorithm, we have that the induced subgraph $H$ on vertices $v_1,v_2,\ldots,v_i$ of $G$ is activated by the threshold function $p=t-q$,
that is the target set consisting vertices of threshold $0$  under threshold function $p$ in $H$ activates all the vertices of $H$. This statement holds due to the observation that when we get to step $5$ we know that vertices $v_1,v_2,\ldots,v_{i-1}$ are activated by the target set $\{v_i,v_{i+1},\ldots,v_{n}\}$, and in step $5$ we update threshold function $q$ accordingly to simulate this activation process, finally in step $6$ we set $q_i=t_i$ thus activating vertex $v_i$.
\EPF

\BTHM\label{approx_of_alg_partial_incentive1}
The approximation ratio of Algorithm II is $ O( \log^{3/2} n \cdot OPT) $.
\ETHM
\BPF
The approximation ratio of the algorithm follows from the fact that when we reach step $5$ we know that in any valid
solution $q$  to the instance ($G$,$t$) there is an index $1 \leq j \leq i$ such that necessarily $q_j \geq 1$. Hence during the run of the algorithm step $5$ can be reached at most $OPT$ times.
On the other hand after finishing step $5$ we have $\sum_{j=1}^{i-1} q_j \leq W$. Furthermore $q_i \leq d_i \leq 2W$ as vertex $v_i$ has at least
$\frac{d_i}{2}$ neighbors in the set $\{v_1,v_2,\ldots,v_{i-1}\}$ or at least
$\frac{d_i}{2}$ neighbors in the set $\{v_{i+1},v_{i+2},\ldots,v_{n}\}$. We conclude that
\begin{equation}\label{main_algbound1} \notag
   \sum_{j=1}^{i} q_j \leq 3W = O( \log^{3/2} n \cdot CW(G)) = O( \log^{3/2} n \cdot OPT)
\end{equation}
Where the last equality follows from Lemma \ref{pit_alg_lemma1}.
Thus the total weight of the partial incentive function returned by the algorithm is $O( \log^{3/2} n \cdot OPT^2)$.
\EPF
Finally we present a slightly different algorithm  for the TSS$_P$($t_v \geq \frac{1}{2}d_v$) problem, with a better approximation ratio for $OPT \gg \sqrt{n}$.
In the algorithm below we assume that we know the weight $OPT$ as we can simply go over all the $O(n^2$) possible weights.
\\ \text{} \\
\fbox{
\begin{minipage}{14.5 cm}
 \textbf{Algorithm III} \\ \text{} \\
 \textbf{Input:} \hspace{3.5pt} A graph $G = (V,E)$ on $n$ vertices and a majority threshold function $t$.\\
 \textbf{Output:} A partial incentive function $q$ of weight at most $\tilde{O}((n \cdot OPT)^{\frac{2}{3}} )$ \\
 (where $OPT$ is the weight of an optimal solution).
 \begin{enumerate}
 \item Set $k=\left\lceil \left(\frac{n^2}{OPT}\right)^\frac{1}{3}\right\rceil$ and set $r=\left\lceil\frac{n}{k}\right\rceil$.
 \item Let $(v_1,v_2,\ldots,v_n)$ be a cutwidth arrangement of graph $G$ of width at most $W = O( \log^{3/2} n \cdot CW(G) )$.
 \item For each $1\leq j \leq r$ do the following: set $q_j = \min\{t_j,n_j\}$ where $n_j$ is the number of neighbors vertex $v_j$ has in the vertex set $\{v_{r+1},v_{r+2},\ldots,v_n\}$ and $t_j = t(v_j)$.
 \item Let $G^{new}$ be the induced subgraph of $G$ on vertices $v_1,v_2,\ldots,v_r$ with the auxiliary threshold function $t^{new}(v_i) = t_i - q_i$.
 \item Apply the algorithm presented in Theorem \ref{simple_pit1} on the instance ($G^{new}$,$t^{new}$) of the TSS$_P$ problem, and let $q^{new}$ be the partial incentive function returned.
 \item For each $1\leq j \leq q$ do the following: set $q_j = q_j + q^{new}_j$ .
 \item Let ($G'$,$t'$) be an instance of the TSS$_P$ problem where graph $G'$ is the induced subgraph of $G$ on vertices $v_{r+1},v_{r+2},\ldots,v_n$  and
  threshold function $t'$ for the vertices of $G'$ is the threshold function resulting from activating vertices $v_1,v_2,\ldots,v_r$ of the instance ($G$,$t$).
\item Renumber the vertices of $G'$ to be $v_1,v_2,\ldots,v_{n-r}$. Set $G=G'$,$t=t'$ and $n=n-r$.
\item If $r > n$ then set $r=n$.
\item If $n>0$ then goto step 2.
 \end{enumerate}
\end{minipage}
}
\\ \text{} \\ \text{} \\
\BTHM\label{approx_of_alg_partial_incentive2}
The approximation ratio of Algorithm III is $\tilde{O}\left(\left(\frac{n^2}{OPT}\right)^\frac{1}{3}\right)$
\ETHM
\BPF
We assume without loss of generality that $k$ divides $n$.
Step $3$ of the algorithm is executed at most $k$ times and  the total contribution of weight to the partial incentive function $q$ from these executions of step $3$  is at most
\begin{equation}\label{step3analysis}
  \tilde{O}(k \cdot CW(G)) = \tilde{O}(k \cdot OPT)
\end{equation}
Where the equality follows from Lemma \ref{pit_alg_lemma1}.
Step $6$ of the algorithm is also executed at most $k$ times and the total contribution of weight to the partial incentive function $q$ from these executions of step $6$  is at most
\begin{equation}\label{step6analysis}
  r \sum_{i=1}^{k} \sqrt{2OPT_i}
\end{equation}
Where $\sum_{i=1}^{k} OPT_i \leq OPT$ and bound (\ref{step6analysis}) follows from Theorem \ref{simple_pit1}. Notice that $OPT_i$ is the weight of the optimal partial incentive function
of the instance $(G^{new}$,$t^{new})$ in the $i$-th execution of step $5$.
Furthermore we have that
\begin{align}\label{step6long}
   r \sum_{i=1}^{k} \sqrt{2OPT_i} &\leq r \sqrt{k \cdot 2OPT} &&\text{by Jensen's Inequality} \notag \\
  &= \frac{n}{\sqrt{k}} \cdot  \sqrt{2OPT} &&\text{as $r=\frac{n}{k}$}  \notag \\
  &=  O\left((n \cdot OPT)^{\frac{2}{3}} \right) &&\text{as $k=\left\lceil \left(\frac{n^2}{OPT}\right)^\frac{1}{3}\right\rceil$ }
\end{align}
And
\begin{align}\label{step3long}
  \tilde{O}(k \cdot OPT) &=  \tilde{O}\left((n \cdot OPT)^{\frac{2}{3}} \right) &&\text{as $k=\left\lceil \left(\frac{n^2}{OPT}\right)^\frac{1}{3}\right\rceil$ }
\end{align}
Hence we have by (\ref{step3long}) and (\ref{step6long}) that the total weight of the partial incentive function $q$ is
\[
 \tilde{O}(k \cdot OPT) + r \sum_{i=1}^{k} \sqrt{2OPT_i} = \tilde{O}\left((n \cdot OPT)^{\frac{2}{3}} \right)
\]
and thus we are done.
\EPF
\BTHM
(Restatement of Theorem \ref{thm:Pweak})
Given an instance of $(G,t)$ of the TSS$_P$($t_v \geq \frac{1}{2}d_v$) problem, where $G$ is a graph on $n$ vertices, an $\tilde{O}(\sqrt{n})$-approximate solution can be found in polynomial time.
\ETHM
\BPF
Apply algorithms $II$ and $III$ on the instance($G$,$t$) and take the solution of least weight.
This gives the required bound as by Theorem \ref{approx_of_alg_partial_incentive1} and Theorem \ref{approx_of_alg_partial_incentive2} we get an approximation ratio of
 $\tilde{O}\left(\min\left(OPT,\left(\frac{n^2}{OPT}\right)^\frac{1}{3}\right)\right) = \tilde{O}(\sqrt{n})$.
\EPF
\BTHM
(Restatement of Theorem \ref{thm:Pstrong})
Given an instance of ($G$,$t$) of the TSS$_P$($t_v > \frac{1}{2}d_v$) problem,
where $G$ is a graph on $n$ vertices, an $\tilde{O}(n^{\frac{1}{3}})$-approximate solution can be found in polynomial time.
\ETHM
\BPF
By Lemma $1$ of \cite{cordasco2018whom} we have
\begin{equation}\label{whomtobefriendpaper2}
OPT \geq \sum_{i=1}^{n} t_i - |E|
\end{equation}
Where $|E|$ is the number of edges in $G$ and $t_i=t(v_i)$. Now as $t$ is a strict majority function we conclude from (\ref{whomtobefriendpaper2}) that
\begin{equation}\label{whomtobefriend3}
OPT \geq \frac{n}{2}
\end{equation}
Hence applying Algorithm III on the instance $(G,t)$ of the TSS$_P$($t_v > \frac{1}{2}d_v$) problem will result in a solution with approximation ratio
\[
\tilde{O}\left(\left(\frac{n^2}{OPT}\right)^\frac{1}{3}\right) = \tilde{O}\left( n^{\frac{1}{3}} \right)
\]
Where the last equality is by applying (\ref{whomtobefriend3}).
\EPF

We mention that using standard reductions one can show $APX$-hardness for the TSS$_P$(${t_v \geq \frac{1}{2}d_v}$) and TSS$_P$($t_v > \frac{1}{2}d_v$) problems.
We sketch below such a proof for the TSS$_P$($t_v \geq \frac{1}{2}d_v$) problem.
\BTHM
The TSS$_P$($t_v \geq \frac{1}{2}d_v$) problem in graphs of maximum degree $6$ is $APX$-hard.
\ETHM
\BPF
(Sketch) We reduce from the cubic vertex cover problem which is $APX$-hard
(\cite{DBLP:conf/ciac/AlimontiK97}). Let $H$ be the cubic graph on $n$ vertices $v_1,v_2,\ldots,v_n$, in which we want to find a minimum vertex cover.
Add to graph $H$ a threshold function $t$ such that for all $1 \leq i \leq n$ we have $t(v_i) = 3$. Hence in the TSS instance $(H,t)$ each vertex has a degree equal to its threshold. Thus graph $H$ has a vertex cover of size $k$ if and only if the TSS instance $(H,t)$ has a target set of size $k$ (this was first observed in \cite{chen1}).

Let $H'$ be the graph constructed from graph $H$ by doing the following for all $1 \leq i \leq n$: add vertices $u^1_i,u^2_i,u^3_i,u^4_i$ such that $u^1_i$ is adjacent to
$u^2_i,u^3_i,u^4_i$, and $v_i$ is adjacent to $u^2_i,u^3_i,u^4_i$. Now set $t(u^1_i) = t(u^2_i) = t(u^3_i) = t(u^4_i) = 1$. In Theorem $1$ of \cite{cordasco2018whom} it is proven that
the instance $(H',t)$ of TSS$_P$ has a partial incentive function of weight $k$ if and only if the TSS instance $(H,t)$ has a target set of size $k$ and thus we conclude that the instance $(H',t)$ of TSS$_P$ has a partial incentive function of weight $k$ if and only if graph $H$ has a vertex cover of size $k$.

Finally we create graph $H''$ in the following manner. Let $H''$ be the graph constructed from graph $H'$ by doing the following for all $1 \leq i \leq n$:
add vertices $q^1_i,q^2_i,q^3_i,q^4_i$ such that $q^1_i$ is adjacent to $q^2_i,q^3_i$, and  $q^4_i$ is adjacent to $q^2_i,q^3_i,u^1_i$.
Set threshold function $t'=t$. Now set $t'(u^1_i) = 2$ (thus increasing it by one unit). Set $t'(q^1_i) = t'(q^2_i) = t'(q^3_i) = 1$ and set $t'(q^4_i)=2$. It is clear that the instance
$(H'',t')$ of TSS$_P$($t_v \geq \frac{1}{2}d_v$) has a partial incentive function of weight $n+k$ if and only if the instance $(H',t)$ of TSS$_P$ has a partial incentive function of weight $k$ and we are done as $k=\Omega(n)$ since $H$ is cubic and hence the size of its vertex cover is $\Omega(n)$.
\EPF
In Section \ref{degeberate_section1} we defined the TSS(degenrate) problem.
We note that this problem in the partial incentives model denoted by  TSS$_P$(degenrate) can be solved in polynomial time.
Now we shall prove Theorem \ref{surprising_polytime1} in two stages. First we prove the following claim.
\BTHM\label{surprising_polytime11}
The  TSS$_P$(degenrate) problem can be solved in polynomial time.
\ETHM
\BPF
(Sketch) Let $(G,t)$ be an instance of the TSS$_P$(degenrate) problem. Where the vertices of graph $G$ are $v_1,\ldots,v_n$ and the thresholds are $t_1,t_2,\ldots,t_n$.
By Lemma $1$ of \cite{cordasco2018whom} we have
\begin{equation}\label{whomtobefriendpaper143}
OPT \geq \sum_{i=1}^{n} t_i - |E|
\end{equation}
Furthermore using the same ideas as in Section \ref{degeberate_section1} we can always find in polynomial time a partial incentive function for the instance of weight exactly $\sum_{i=1}^{n} t_i - |E|$ which is optimal by (\ref{whomtobefriendpaper143}).
\EPF
\BCR
The TSS$_P$($t_v \in \{d_v -1 , d_v\})$) problem and the TSS$_P$($t_v \in \{1 , d_v\})$) problem can be solved in polynomial time.
\ECR
\BPF
(Sketch) This can be proven by converting an instance of such a problem to an instance of the TSS$_P$(degenrate) problem.
See Appendix \ref{TSSP_polycases2}.
\EPF
We note that the TSS($t_v \in \{d_v -1 , d_v\})$) problem and the TSS($t_v \in \{1 , d_v\})$) problem are $NP$-hard.

\section*{Acknowledgements}
Work supported in part by the Israel Science Foundation (grant No. 1388/16).

\bibliographystyle{alpha}

\newcommand{\etalchar}[1]{$^{#1}$}

\begin{appendix}
\section{Hardness of the directed feedback vertex set problem on restricted graphs}\label{appendix1DFVS}
We call a directed graph $G$ restricted if for each vertex $v \in V(G)$ one of the following two conditions holds.
\begin{enumerate}
  \item $d^-(v)=1$ and $d^+(v)=2$.
  \item $d^+(v)=1$ and $d^-(v)=2$.
\end{enumerate}
and furthermore the number of vertices of indegree $1$ in $G$ equals the number of vertices of outdegree $1$ in $G$.
In this section we shall prove Theorem \ref{restrictedgraphtheorem0} which states that it is $NP$-hard to approximate the DFVS problem on restricted directed graphs within any
constant factor assuming the $UGC$.
We start with the following Lemma.
\BL\label{restrictedgraphlem1}
 It is $NP$-hard to approximate the DFVS problem on directed graphs where each vertex has indegree at least $2$ and outdegree at least $2$, within any
constant factor assuming the $UGC$.
\EL
\BPF
(Sketch)  This follows by a reduction from the DFVS problem in general graphs. The reduction is by the contraction operations stated in Proposition $5.1$ of \cite{DBLP:journals/jal/LevyL88}.
\EPF
Now we shall sketch a proof of Theorem \ref{restrictedgraphtheorem0} by a reduction from the DFVS problem on directed graphs where each vertex has indegree at least $2$ and outdegree at least $2$.
Let $G$ be an instance of the DFVS problem on directed graphs where each vertex has indegree at least $2$ and outdegree at least $2$.
Assume that graph $G$ has $n$ vertices.
For each vertex $v \in V(G)$ we will replace vertex $v$ with the following gadget:
Let $u_1,u_2,\ldots,u_k$ be the in-neighbors of vertex $v$ in $G$ (where $k=d^-(v) \geq 2$).
Let $w_1,w_2,\ldots,w_l$ be the out-neighbors of vertex $v$ in $G$ (where $l=d^+(v) \geq 2$). \\
Now we remove vertex $v$ from graph $G$ and add vertices $u'_1,u'_2,\ldots,u'_{k-1}$ and $w'_1,w'_2,\ldots,w'_{l-1}$ to graph $G$.
Now we add directed edges to graph $G$ in the following manner.
\begin{itemize}
  \item For all $1 \leq i \leq k-2$ we add a directed edge $(u'_i,u'_{i+1})$ (that is $u'_i$ is an in-neighbor of $u'_{i+1}$).
  \item For all $1 \leq i \leq l-2$ we add a directed edge $(w'_i,w'_{i+1})$.
  \item We add a directed edge $(u'_{k-1},w'_1)$.
  \item For all $1 \leq i \leq k-1$ we add a directed edge $(u_{i+1},u'_i)$.
  \item We add a directed edge $(u_1,u'_1)$.
  \item For all $1 \leq i \leq l-1$ we add a directed edge $(w'_i,w_i)$.
  \item We add a directed edge $(w'_{l-1},w_l)$.
\end{itemize}
Notice that the induced subgraph of $G$ consisting of vertices $u'_1,u'_2,\ldots,u'_{k-1}$ and $w'_1,w'_2,\ldots,w'_{l-1}$ is a directed path.
Denote the resulting graph by $G'$ and denote by $G$ the original graph (before the replacement operations).
Observe that we have the following.
\begin{itemize}
  \item The number of vertices of indegree $1$ in $G'$ equals to the sum of outdegrees of graph $G$ minus $n$.
  \item The number of vertices of outdegree $1$ in $G'$ equals to the sum of indegrees of graph $G$ minus $n$.
\end{itemize}
Now as in any directed graph the sum of indegrees equals the sum of outdegrees we conclude that the
number of vertices of indegree $1$ in $G'$ equals the number of vertices of outdegree $1$ in $G'$.
Notice that graph $G$ has a feedback vertex set of size at most $k$ if and only if graph $G'$ has a feedback vertex set of size at most $k$.
\section{Polynomial subcases of TSS$_P$}\label{TSSP_polycases2}
\BL\label{tiny_fact_tss_p1}
Let $G = (V,E)$ be a connected graph on $n$ vertices and let $t$ be a threshold function
$t: V(G) \rightarrow \mathbb{N}$ such that $t(v) \geq d(v)-1$ for all $v \in V(G)$, furthermore assume that there is a vertex $v \in V(G)$ for which $t(v) \geq d(v)$. Then
function $t$ is a degenerate threshold function.
\EL
\BPF
The proof is by induction on the number of vertices of graph $G$.
It holds trivially for graphs on one or two vertices. Now as graph $G$ has a vertex $v \in V(G)$ for which $t(v) \geq d(v)$ we can remove this vertex and in the resulting graph
each connected component $G'$ will contain a vertex $v'$ such that $t(v') \geq d_{G}(v')-1 = d_{G'}(v')$ (where $v'$ is a neighbor of $v$ in $G$) and we are done by the induction hypothesis.
\EPF
Recall that in Theorem \ref{surprising_polytime11} we have shown that the  TSS$_P$(degenrate) problem can be solved in polynomial time.
We will use this theorem to prove the following.
\BTHM
The TSS$_P$($t_v \in \{d_v -1 , d_v\})$) problem can be solved in polynomial time.
\ETHM
\BPF
Let $G$ be a connected graph on $n$ vertices $v_1,v_2,\ldots,v_n$ , degrees $d_1,d_2,\ldots,d_n$ and thresholds $t_1,t_2,\ldots,t_n$ , where for all $1 \leq i \leq n$ we have $t_i \geq d_i-1$.
If there is an $i$ for which $t_i \geq d_i$ then threshold function $t$ is a degenerate threshold function by Lemma \ref{tiny_fact_tss_p1} and we are done by Theorem \ref{surprising_polytime11}.
Otherwise we have  for all $1 \leq i \leq n$ that $t_i = d_i-1$. This means that there is an edge of $G$ that does not participate in the activation process (in particular an edge which touches the last vertex to be activated in the activation process as its threshold is smaller than its degree). We guess which edge it is and remove it from $G$ and call the resulting graph $G'$.
By Lemma \ref{tiny_fact_tss_p1} threshold function $t$ is a degenerate threshold function for each connected component of graph $G'$ and thus we are done by  Theorem \ref{surprising_polytime11}.
\EPF
\BTHM
The TSS$_P$($t_v \in \{1 , d_v\})$) problem can be solved in polynomial time.
\ETHM
\BPF
(Sketch)
Let $G$ be graph such that for each vertex $v \in G$ we have $t(v) = 1$ or $t(v) = d(v)$.
We partition the vertices of $G$ into two sets $A$ and $B$ such that a vertex $v \in B$ if and only if $t(v) = 1$ and $A = V(G) \setminus B$.
Construct from graph $G$ a multigraph $G'$ in the following manner: For each  connected component $Q$ of graph $G[B]$ we contract the vertices of $Q$ into a single vertex $q$ with threshold $1$ (another way to look at it is that we remove the vertices of $Q$ from the graph and add a vertex $q$ of threshold $1$ and we connect
each vertex $v \in A$ to vertex $q$ by $k$ parallel edges if vertex $v$ had $k$ neighbors in $Q$). Let $Q'$ be the set of all vertices created in the contraction step above. Notice that $Q'$ is an independent set in graph $G'$.
Now we convert multigraph $G'$ into a simple graph $H$ by replacing each edge $(u,v)$ of $G'$ by a new vertex $d$ of threshold $1$ and edges $(u,d)$ and $(v,d)$,
we denote the set of vertices added in this process by $D$. Hence the vertices of graph $H$ can be partitioned into the following three sets.
\begin{enumerate}
  \item The set $A$ as defined above in which for all $v \in A$ we have $t(v) = d(v)$.
  \item The set $D$ as defined above in which all vertices are of degree $2$ and threshold $1$ and each vertex in $D$ has a neighbor in $A$.
  \item The set $Q'$ as defined above in which all vertices are of threshold $1$ and the vertices of $Q'$ induce an independent set in graph $H$.
\end{enumerate}
Now notice that the threshold function $t$ for the graph $H$ is a degenerate threshold function
(the degeneracy order would be first all the vertices in $Q'$, then all the vertices in $D$ and finally all the vertices in $A$).
Thus we are done by  Theorem \ref{surprising_polytime11}.
\EPF
\end{appendix}\

\end{document}